\documentclass[a4paper,12pt]{article}
\usepackage{graphicx}
\usepackage{amsmath}
\usepackage{latexsym}
\usepackage{setspace}
\usepackage{color}
\usepackage[margin=1in]{geometry}
\usepackage{times}
\usepackage{txfonts}
\usepackage{mathptmx}
\usepackage{listings}
\usepackage{subfloat}
\usepackage{subfig}
\usepackage{authblk}
\usepackage[comma,numbers,round]{natbib} 
\makeatletter
\renewcommand\@biblabel[1]{#1.} 
\makeatother

\def\cd{{\rm CD8}^{+}}

\usepackage{color} 
\definecolor{mygreen}{RGB}{28,172,0} 
\definecolor{mylilas}{RGB}{170,55,241}
\title{\huge{On the role of $\cd$ T cells in determining recovery time from influenza virus infection}}

\author[1]{Pengxing Cao}
\author[2,3]{Zhongfang Wang}
\author[1]{Ada W. C. Yan}
\author[4,5]{Jodie McVernon}
\author[3]{Jianqing Xu}
\author[6]{Jane M. Heffernan}
\author[2]{Katherine Kedzierska}
\author[1,4,5]{James M. McCaw\thanks{Correspondence: jamesm@unimelb.edu.au}}
\affil[1]{School of Mathematics and Statistics, The University of Melbourne, Melbourne, Australia.}
\affil[2]{Department of Microbiology and Immunology, University of Melbourne, at the Peter Doherty Institute for Infection and Immunity, Parkville, Victoria, Australia}
\affil[3]{Shanghai Public Health Clinical Center and Institutes of Biomedical Sciences, Key Laboratory of Medical Molecular Virology of Ministry of Education/Health, Shanghai Medical College, Fudan University, Shanghai, China}
\affil[4]{Centre for Epidemiology and Biostatistics, Melbourne School of Population and Global Health, The University of Melbourne, Melbourne, Australia.}
\affil[5]{Modelling and Simulation, Infection and Immunity Theme, Murdoch Childrens Research Institute, The Royal Children's Hospital, Parkville, Victoria, Australia.}
\affil[6]{Modelling Infection and Immunity Lab, Centre for Disease Modelling, York Institute for Health Research, York University, Toronto, Ontario, Canada.}

\begin{document}

\date{}
\maketitle
\vspace{1.5cm}

\thispagestyle{empty}

\newpage

\section*{Abstract}
Myriad experiments have identified an important role for $\cd$ T cell response mechanisms in determining recovery from influenza A virus infection. Animal models of influenza infection further implicate multiple elements of the immune response in defining the dynamical characteristics of viral infection. To date, influenza virus models, while capturing particular aspects of the natural infection history, have been unable to reproduce the full gamut of observed viral kinetic behaviour in a single coherent framework. Here, we introduce a mathematical model of influenza viral dynamics incorporating all major immune components (innate, humoral and cellular) and explore its properties with a particular emphasis on the role of cellular immunity. Calibrated against a range of murine data, our model is capable of recapitulating observed viral kinetics from a multitude of experiments. Importantly, the model predicts a robust exponential relationship between the level of effector $\cd$ T cells and recovery time, whereby recovery time rapidly decreases to a fixed minimum recovery time with an increasing level of effector $\cd$ T cells. We find support for this relationship in recent clinical data from influenza A(H7N9) hospitalised patients. The exponential relationship implies that people with a lower level of naive $\cd$ T cells may receive significantly more benefit from induction of additional effector $\cd$ T cells arising from immunological memory, itself established through either previous viral infection or T cell-based vaccines. 

\newpage
\section*{Introduction}

Invasion of influenza virus into a host's upper respiratory tract leads to infection of healthy epithelial cells and subsequent production of progeny virions \cite{Taubenbergeretal2008}. Infection also triggers a variety of immune responses. In the early stage of infection a temporary non-specific response (innate immunity) contributes to the rapid control of viral growth while in the late stage of infection, the adaptive immune response dominates viral clearance \cite{Kreijtzetal2011}. The early immune response involves production of antiviral cytokines and cells, e.g. type 1 interferon (IFN) and natural killer cells (NK cells), and is independent of virus type \cite{Goodbournetal2000,Sadleretal2008,Bironetal1999,Jostetal2013,Iwasakietal2014}. In the special case of a first infection in a naive host, the adaptive immune response, mediated by the differentiation of naive T cells and B cells and subsequent production of virus-specific T cells and antibodies \cite{Muralietal1998,Kreijtzetal2011}, leads to not only a prolonged killing of infected cells and virus but also the formation of memory cells which can generate a rapid immune response to secondary infection with the same virus \cite{Wherryetal2004,LaGrutaetal2014}.  

$\cd$ T cells, which form a major component of adaptive immunity, play an important role in efficient viral clearance \cite{Zhangetal2011}.  However, available evidence suggests they are unable to clear virus in the absence of antibodies \cite{Iwasakietal1977,Fangetal2005} except in hosts with a very high level of pre-existing naive or memory $\cd$ T cells \cite{Grahametal1997,Moskophidisetal1998,Valkenburgetal2010}. Some studies indicate that depletion of $\cd$ T cells could decrease the viral clearance rate and thus prolong the duration of infection \cite{Yapetal1978,Wellsetal1981,Houetal1992,Benderetal1992}. Furthermore, a recent study of human A(H7N9) hospitalized patients has implicated the number of effector $\cd$ T cells as an important driver of the duration of infection \cite{Wangetal2015}. This diverse experimental and clinical data, sourced from a number of host-species, indicates that timely activation and elevation of $\cd$ T cell levels may play a major role in the rapid and successful clearance of influenza virus from the host. These observations motivate our modeling study of the role of $\cd$ T cells in influenza virus clearance.

Viral dynamics models have been extensively applied to the investigation of the antiviral mechanisms of $\cd$ T cell immunity against a range of pathogens, with major contributions for chronic infections such as HIV/SIV \cite{Perelsonetal1996,Perelsonetal2002,Antiaetal2003,Chaoetal2004,Hrabaetal1995,DeBoeretal2007}, HTLV-I \cite{Limetal2014} and chronic LCMV \cite{Althausetal2007,Leetal2015}. However, for acute infections such as measles \cite{Heffernanetal2008} and influenza \cite{Bocharovetal1994,Changetal2007,Hanciogluetal2007,Handeletal2008,Leeetal2009,Saenzetal2010,Miaoetal2010,Dobrovolnyetal2013,Reperantetal2014,Crausteetal2015,Zarnitsynaetal2016}, highly dynamical interactions between the viral load and the immune response occur within a very short time window, presenting new challenges for the development of models incorporating $\cd$ T cell immunity.  

Existing influenza viral dynamics models, introduced to study specific aspects of influenza infection, are limited in their ability to capture all major aspects of the natural history of infection, hindering their use in studying the role of $\cd$ T cells in viral clearance. Some models show a severe depletion of target cells (i.e healthy epithelial cells susceptible to viral infection) after viral infection \cite{Hanciogluetal2007,Leeetal2009,Saenzetal2010,Miaoetal2010,Reperantetal2014}. Depletion may be due to either infection or immune-mediated protection. Either way, these models are arguably incompatible with recent evidence that the host is susceptible to re-infection with a second strain of influenza a short period following primary exposure \cite{Laurieetal2015}. Furthermore, as reviewed by Dobrovolny \emph{et al.} \cite{Dobrovolnyetal2013}, target cell depletion in these models strongly limits viral expansion so that virus can be effectively controlled or cleared at early stage of infection even in the absence of adaptive immunity, which contradicts the experimental finding that influenza virus remains elevated in the absence of adaptive immune response \cite{Krisetal1988}. While a few models do avoid target cell depletion \cite{Bocharovetal1994,Changetal2007}, they either assume immediate replenishment of target cells \cite{Bocharovetal1994} or a slow rate of virus invasion into target cells resulting in a much delayed peak of virus titer at day 5 post-infection (rather than the observed peak at day 2) \cite{Changetal2007}. Moreover, models with missing or unspecified major immune components, e.g. no innate immunity \cite{Antiaetal2003,Chaoetal2004,Leeetal2009,Miaoetal2010}, no antibodies \cite{Antiaetal2003,Chaoetal2004,Changetal2007,Crausteetal2015,Zarnitsynaetal2016} or unspecified adaptive immunity \cite{Reperantetal2014}, also indicate the need for further model development. For an in-depth review of the current virus dynamics literature on influenza, we refer the reader to the excellent article by Dobrovolny \emph{et al.} \cite{Dobrovolnyetal2013}. 

\begin{figure}[ht!]
\centering
\includegraphics[scale=0.6]{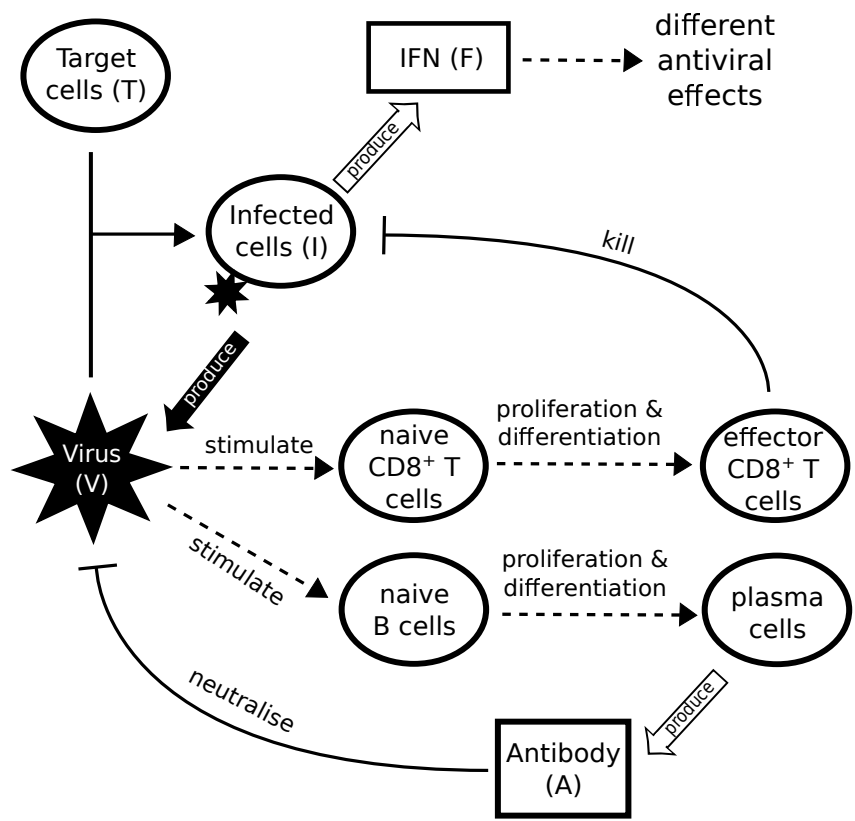}
\caption{Schematic diagram showing the major components of viral infection and the immune response. Infection starts when virus binds to healthy epithelial cells (target cells). Infected cells release new virus and produce cytokines such as IFN. IFN is a major driver of innate immunity, responsible for effective control of rapid viral growth and expansion. Virus further stimulates naive $\cd$ T cells and B cells to produce effector $\cd$ T cells and antibodies, responsible for final clearance of virus.}
\end{figure}

In this paper, we construct a within-host model of influenza viral dynamics in naive (i.e.\ previously unexposed) hosts that incorporates the major components of both innate and adaptive immunity and use it to investigate the role of $\cd$ T cells in influenza viral clearance. The model is calibrated against a set of published murine data from Miao \emph{et al.} \cite{Miaoetal2010} and is then validated through demonstration of its ability to qualitatively reproduce a range of published data from immune-knockout experiments \cite{Iwasakietal1977,Yapetal1978,Wellsetal1981,Krisetal1988,Fangetal2005,Miaoetal2010}. Using the model, we find that the recovery time --- defined to be the time when virus titer first drops below a chosen threshold in the (deterministic) model --- is negatively correlated with the level of effector $\cd$ T cells in an approximately exponential manner. To the best of our knowledge, this relationship, with support in both H3N2-infected mice and H7N9-infected humans \cite{Wangetal2015}, has not been previously identified. The exponential relationship between $\cd$ T cell level and recovery time is shown to be remarkably robust to variation in a number of key parameters, such as viral production rate, IFN production rate, delay of effector $\cd$ T cell production and the level of antibodies. Moreover, using the model, we predict that people with a lower level of naive $\cd$ T cells may receive significantly more benefit from induction of additional effector $\cd$ T cells. Such production, arising from immunological memory, may be established through either previous viral infection or T cell-based vaccines.

\section*{Methods}

\subsection*{The model}
The model of primary viral infection is a coupled system of ordinary and delay differential equations, consisting of three major components (see Fig.\ 1 for a schematic diagram). Eqs.\ 1--3 describe the process of infection of target cells by influenza virus and are a major component in almost all models of virus dynamics in the literature. Eqs.\ 4 and 5 model IFN-mediated innate immunity \cite{Paweleketal2012,Caoetal2015}. Thirdly, adaptive immunity including $\cd$ T cells and B cell-produced antibodies for killing infected cells and neutralizing influenza virus respectively are described by Eqs.\ 6--11.  

\begin{align}
\frac{dV}{dt} & =p_VI-\delta_VV-\kappa_{S} VA_S-\kappa_{L} VA_L-\beta VT, \\
\frac{dT}{dt} & =g_T(T+R)(1-\frac{T+R+I}{T_0})-\beta' VT+\rho R-\phi FT, \\
\frac{dI}{dt} & =\beta' VT-\delta_I I-\kappa_{N} IF-\kappa_{E} IE, \\
\frac{dF}{dt} & =p_FI-\delta_FF, \\
\frac{dR}{dt} & =\phi FT-\rho R, \\
\frac{dC_{n}}{dt} & =-\beta_{Cn}(\frac{V}{V+h_C})C_{n}, \\
\frac{dE}{dt} & = \beta_{Cn}(\frac{V(t-\tau_C)}{V(t-\tau_C)+h_C})C_{n}(t-\tau_C)e^{(p_C\tau_C)} - \delta_EE,\\
\frac{dB_n}{dt} & =-\beta_{Bn}(\frac{V}{V+h_B})B_{n}, \\
\frac{dP}{dt} & = \beta_{Bn}(\frac{V(t-\tau_B)}{V(t-\tau_B)+h_B})B_{n}(t-\tau_B)e^{(p_B\tau_B)} - \delta_PP,\\
\frac{dA_S}{dt} & =p_SP-\delta_SA_S, \\
\frac{dA_L}{dt} & =p_LP-\delta_LA_L.
\end{align}

In further detail, Eq.\ 1 indicates that the change in viral load ($V$) is controlled by four factors: the production term ($p_VI$) in which virions are produced by infected cells ($I$) at a rate $p_V$ \cite{Baccametal2006,Saenzetal2010,Paweleketal2012}; the viral natural decay/clearance ($\delta_VV$) with a decay rate of $\delta_V$; the viral neutralisation terms ($\kappa_S VA_S$ and $\kappa_L VA_L$) by antibodies (both a short-lived antibody response $A_S$ driven by, e.g. IgM, and a longer-lived antibody response $A_L$ driven by, e.g. IgG and IgA \cite{Iwasakietal1977,Miaoetal2010}), and a consumption term ($\beta VT$) due to binding to and infection of target cells ($T$). In Eq.\ 2, the term $g_T(T+R)(1-(T+R+I)/T_0)$ models logistic regrowth of the target cell pool \cite{Caoetal2015}. Both target cells ($T$) and resistant cells ($R$, those protected due to IFN-induced antiviral effect) can produce new target cells, with a net growth rate proportional to the severity of infection, $1-(T+R+I)/T_0$ (i.e. the fraction of dead cells). $T_0$ is the initial number of target cells and the maximum value for the target cell pool \cite{Hanciogluetal2007}. Target cells ($T$) are consumed by virus ($V$) due to binding ($\beta' VT$), the same process as $\beta VT$. Note that $\beta$ and $\beta'$ have different measurement units due to different units for viral load ($V$) and infected cells ($I$). As already mentioned, the innate response may trigger target cells ($T$) to become resistant ($R$) to virus, at rate $\phi FT$. Resistant cells lose protection at a rate $\rho$ \cite{Paweleketal2012}. This process also governs the evolution of virus-resistant cells ($R$) in Eq.\ 5.

Eq.\ 3 describes the change of infected cells ($I$). They increase due to the infection of target cells by virus ($\beta' VT$) and die at a (basal) rate $\delta_I$. Two components of the immune response increase the rate of killing of infected cells. IFN-activated NK cells kill infected cells at a rate $\kappa_N IF$ \cite{Hwangetal2012,Jostetal2013,Paweleketal2012,Caoetal2015}. Effector $\cd$ T cells ($E$) --- produced through differentiation from naive $\cd$ T cells $C_n$ in Eq.\ 6 --- kill at a rate $\kappa_E IE$. Of note our previous work has demonstrated that models of the innate response containing only IFN-induced resistance for target cells (state $R$; Eq.\ 5), while able to maintain a population of healthy uninfected cells, still control viral kinetics through target cell depletion, and therefore cannot reproduce viral re-exposure data \cite{Caoetal2015,Laurieetal2015}. Given our interest in analysing a model that prevents target cell depletion, inclusion of IFN-activated NK cells (term $\kappa_N IF$) is an essential part of the model construction. 

Eq.\ 4 models the innate response, as mediated by IFN ($F$). IFN is produced by infected cells at a rate $p_F$ and decays at a rate $\delta_F$ \cite{Caoetal2015}. 

Eq.\ 6 models stimulation of naive $\cd$ T cells ($C_n$) into the proliferation/differentiation process by virus at a rate $\beta_{Cn}V/(V+h_C)$), where $\beta_{Cn}$ is the maximum stimulation rate and $h_C$ indicates the viral load ($V$) at which half of the stimulation rate is achieved. Note that this formulation does not capture the process of antigen presentation and $\cd$ T cell activation, but rather is a simple way to establish the essential coupling between the viral load and the rate of $\cd$ T cell activation in the model \cite{Kaechetal2001}. In Eq.\ 7, the production of effector $\cd$ T cells ($E$) is assumed to be an ``advection flux'' induced by a delayed virus-stimulation of naive $\cd$ T cells (the first term on the righthand side of Eq.\ 7). The delayed variables, $V(t-\tau_C)$ and $C_n(t-\tau_C)$, equal zero when $t < \tau_C$. The introduction of the delay $\tau_C$ is to phenomenologically model the delay induced by both naive $\cd$ T cell proliferation/differentiation and effector $\cd$ T cell migration and localization to the site of infection for antiviral action \cite{Cerwenkaetal1999,Lawrenceetal2005,Zarnitsynaetal2016}. The delay also captures the experimental finding that naive $\cd$ T cells continue to differentiate into effector T cells in the absence of ongoing antigenic stimulation \cite{Kaechetal2001,Stipdonketal2001}. The multiplication factor $e^{p_C\tau_C}$ indicates the number of effector $\cd$ T cells produced from one naive $\cd$ T cell, where $p_C$ is the average effector $\cd$ T cell production rate over the delay period $\tau_C$. The exponential form of the multiplication factor is derived based on the assumption that cell differentiation and proliferation follows a first-order advection--reaction equation. Effector $\cd$ T cells decay at a rate $\delta_E$.

Similar to $\cd$ T cells, Eqs.\ 8 and 9 model the proliferation/differentiation of naive B cells, stimulated by virus presentation at rate $\beta_{Bn}V/(V+h_B)$. Stimulation subsequently leads to production of plasma B cells ($P$) after a delay $\tau_B$. The multiplication factor $e^{p_B\tau_B}$ indicates the number of plasma B cells produced from one naive B cell, where $p_B$ is the production rate. Plasma B cells secrete antibodies, which exhibit two types of profiles in terms of experimental observation: a short-lived profile (e.g. IgM lasting from about day 5 to day 20 post-infection) and a longer-lived profile (e.g. IgG and IgA lasting weeks to months) \cite{Iwasakietal1977,Miaoetal2010}. These two antibody responses are modeled by Eqs.\ 10 and 11 wherein different rates of production ($P_S$ and $P_L$) and consumption ($\delta_S$ and $\delta_L$) are assumed. 

\subsection*{Model parameters and simulation}

The model contains 11 equations and 30 parameters (see Table 1). This represents a serious challenge in terms of parameter estimation, and clearly prevents a straightforward application of standard statistical techniques. To reduce uncertainty, a number of parameters were taken directly from the literature, as per the citations in Table.\ 1. The rest were estimated (as indicated in Table\ 1) by calibrating the model against the published data from Miao \emph{et al.} \cite{Miaoetal2010} who measured viral titer, $\cd$ T cell counts and IgM and IgG antibodies in laboratory mice (exhibiting a full immune response) over time during primary influenza H3N2 virus infection (see \cite{Miaoetal2010} for a detailed description of the experiment). The approach to estimating the parameters based on Miao \emph{et al.}'s data is provided in the \emph{Supplementary Material} and the estimated parameter values are given in Table\ 1. Note that the data were presented in scatter plots in the original paper \cite{Miaoetal2010}, while we presented the data here in Mean $\pm$ SD at each data collection time point for a direct comparison with our mean-field mathematical model. 

For model simulation, the initial condition is set to be $(V,T,I,F,R,C_n,E,B_n,P,A_S,A_L) = (V_0,T_0,0,0,0,100,0,100,0,0,0)$ unless otherwise specified. The initial target cell number ($T_0$) was estimated by Petrie \emph{et al.} \cite{Petrieetal2013}. We estimate that of order 100 cells (resident in the spleen) are able to respond to viral infection ($C_n$) (personal communication, N.\ LaGruta, Monash University, Australia). Note that 100 naive $\cd$ T cells might underestimate the actual number of naive precursors that could respond to all the epitopes contained within the virus but does not qualitatively alter the model dynamics and predictions (see \emph{Results} where the naive $\cd$ T cell number is varied between 0 to 200). In the absence of further data, we also use this value for the initial naive B cell number ($B_n$), but again this choice does not qualitatively alter the model predictions. The numerical method and code (implemented in MATLAB, version R2014b, the MathWorks, Natick, MA) for solving the model are provided in the \emph{Supplementary Material}.

\begin{figure}[ht!]
\centering
\includegraphics[scale=0.8]{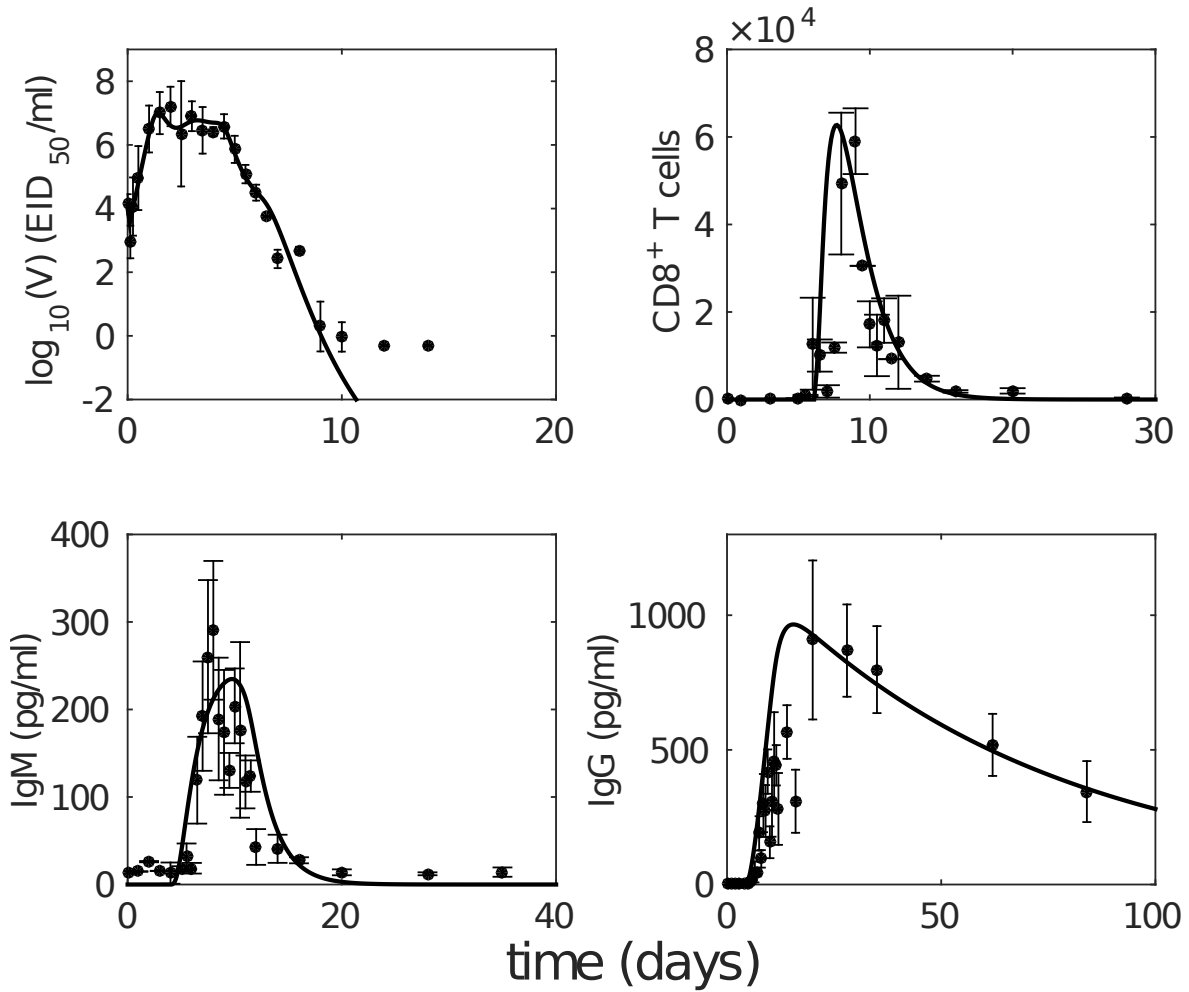}
\caption{The model with estimated parameters (solid curves) captures the murine data from the paper by Miao \emph{et al.}\cite{Miaoetal2010}. The data is shown with error bars ($\rm Mean \pm SD$). Note that due to the limit of detection for the viral load (occurring after 10 days post-infection as seen in viral load data), the last three data points in the upper-left panel were not taken into consideration for model fitting.}
\end{figure}

\subsection*{Analysis of clinical influenza A(H7N9) data}

Clinical influenza A(H7N9) patient data was used to test our model predictions on the relationship between $\cd$ T cell number and recovery time. The data was collected from 12 surviving patients infected with H7N9 virus during the first wave of infection in China in 2013 (raw data is provided in \emph{Dataset S1}; see the paper of Wang \emph{et al.} \cite{Wangetal2015} for details of data collection; this study was reviewed and approved by the SHAPHC Ethics Committee). Note that the clinical data were scarce for some patients. For those patients, we have assumed that the available data are representative of the unobserved values in the neighboring time period. For each patient, we took the average $\rm{IFN\gamma^+}$ $\cd$ T cell number in $10^6$ peripheral blood mononuclear cells (PBMC) for the period from day 8 to day 22 (or the recovery day if it comes earlier) post-admission as a measure of the effector $\cd$ T cell level. This period was chosen \emph{a priori} as it roughly matches the duration of the $\cd$ T cell profile and clinical samples were frequently collected in this period. The average $\cd$ T cell count was given by the ratio of the total area under the data points (using trapezoidal integration) to the number of days from day 8 to day 22 (or the recovery day if it comes earlier). For those patients for which samples at days 8 and/or 22 were missing we specified the average $\cd$ T cell level at the missing time point to be equal to the value from the nearest sampled time available.

\section*{Results}

\subsection*{Model properties and reproduction of published experimental data}

\begin{figure}[ht!]
\centering
\includegraphics[scale=0.9]{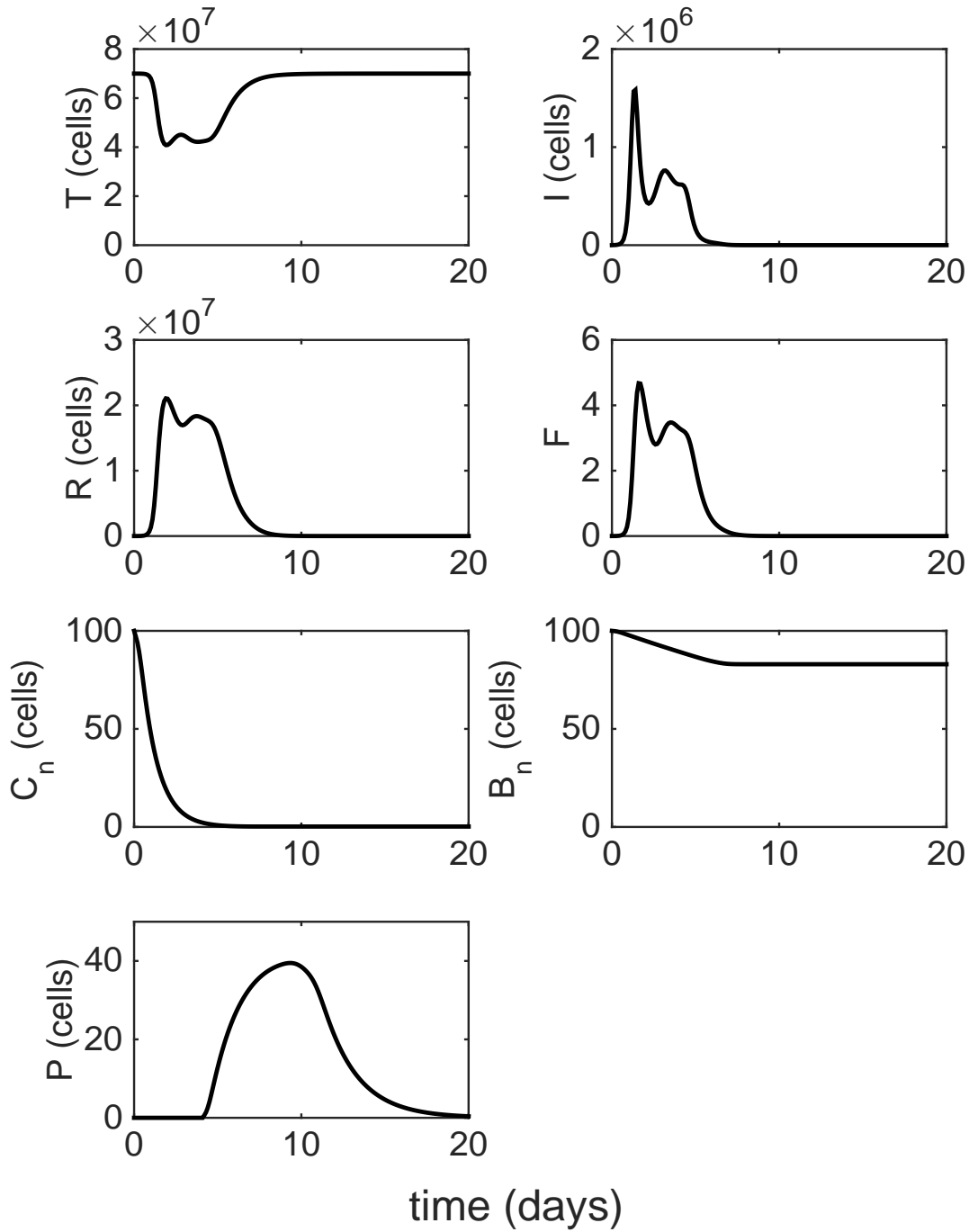}
\caption{Model solution with parameters given in Table 1. Time courses of viral load ($V$), effector $\cd$ T cells ($E$), short-lived antibody response ($A_S$) and long-lived antibody response ($A_L$) have been shown in Fig.\ 2.}
\end{figure}

We first analyze the model behavior in order to establish a clear understanding of the model dynamics. Fig.\ 2 shows solutions (time-series) for the model compartments (viral load, $\cd$ T cells and IgM and IgG antibody) calibrated against the murine data from \cite{Miaoetal2010}. Solutions for the remaining model compartments are shown in Fig.\ 3. The model (with both innate and adaptive components active) prevents the depletion of target cells (see Fig.\ 3 wherein over 50\% of target cells remain during infection) and results in a minor loss of just 10--20\% of healthy epithelial cells (i.e.\ the sum of target cells ($T$) and virus-resistant cells ($R$); see Supplementary Fig.\ S1). The primary driver for the maintenance of the target cell pool during acute viral infection is a timely activation of the innate immune response, and in particular the natural killer cells (Supplementary Fig.\ S2). Since we have previously shown that the target-cell limited model (even with the resistant cell compartment) is unable to reproduce observations from heterologous re-exposure experiments \cite{Laurieetal2015, Caoetal2015}, our model improves upon previous models where viral clearance was only achieved through depletion of target cells (a typical solution shown in Supplementary Fig.\ S2b). Importantly, our result is distinguished from that of Saenz et.\ al.\ \cite{Saenzetal2010}, wherein the healthy cell population was similarly maintained, but primarily through induction of the virus-resistant state, thereby rendering that model incapable of capturing re-infection behavior as established in animal models \cite{Laurieetal2015}.

The modeled viral dynamics exhibits three phases, each dominated by the involvement of different elements of the immune responses (Fig.\ 4). Immediately following infection (0--2 days post-infection) and prior to the activation of the innate (and adaptive) immune responses, virus undergoes a rapid exponential growth (Fig.\ 4a). In the second phase (2--5 days post-infection), the innate immune response successfully limits viral growth (Fig.\ 4a). In the third phase (4--6 days post-infection), adaptive immunity (antibodies and $\cd$ T cells) is activated and viral load decreases rapidly, achieving clearance. Figs.\ 4b and 4c demonstrate the dominance of the different immune mechanisms at different phases. In Fig.\ 4b models with and without immunity are indistinguishable until day 2 (shaded region), before diverging dramatically when the innate and then adaptive immune responses influence the dynamics. In Fig.\ 4c, models with and without an adaptive response only diverge at around day 4 as the adaptive response becomes active. We have further shown that this three-phase property is a robust feature of the model, emergent from its mathematical structure and not a property of fine tuning of parameters (see Supplementary Fig.\ S3). Importantly, it clearly dissects the periods and effect of innate immunity, extending on previous studies of viral infection phases where the innate immune response was either ambiguous or ignored \cite{Iwasakietal1977,Smithetal2010,Miaoetal2010}. 

\begin{figure}[ht!]
\centering
\includegraphics[scale=0.8]{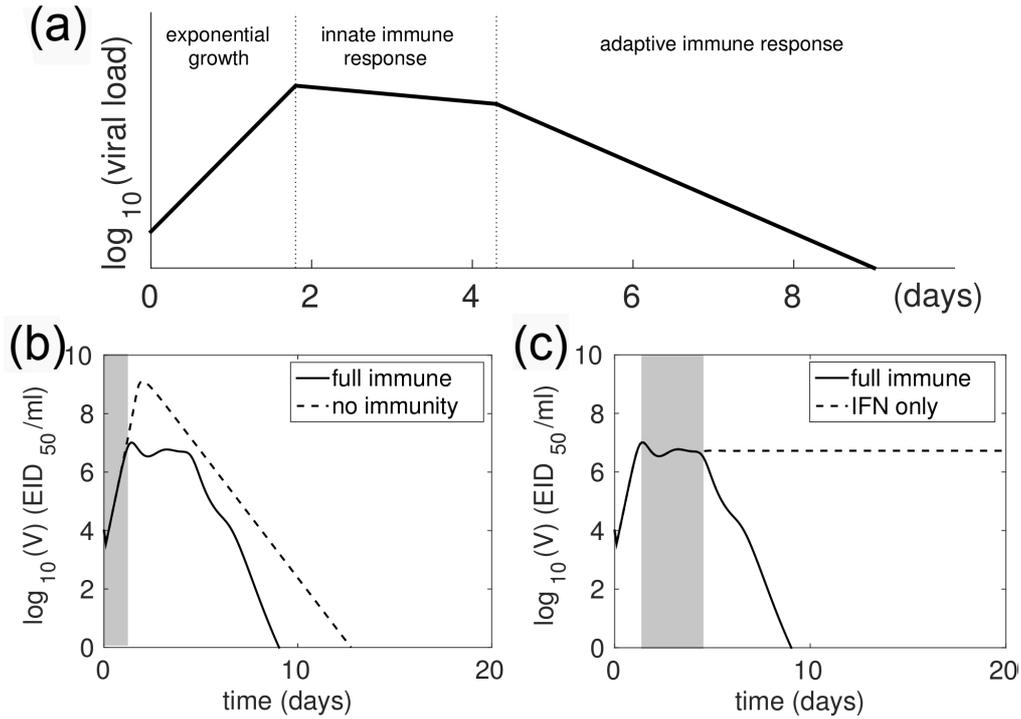}
\caption{The model solution exhibits three-phase behavior following influenza virus infection. Panel (a) schematically represents these three-phases of behavior based on involvement of immune responses. Following infection, virus first undergoes rapid exponential growth before the innate immune response is activated (on day 2).  Innate immunity controls rapid viral expansion to form a plateau in viral load. Adaptive immunity starts to take effect 4--6 days post-infection and is responsible for viral clearance. Panels (b) and (c) demonstrate that the model dynamics follow the three-phase theory. Viral kinetics in the presence of a full immune response is shown by the solid line (in (b) and (c)). In panel (b), the dashed line shows viral kinetics in the complete absence of immunity (innate and adaptive; by letting $p_F = 0$ and $\beta_{Cn} = \beta_{Bn} = 0$ in the model). The trajectories overlap prior to the activation of the innate response, before diverging due to target cell depletion. The shaded region highlights the first phase (exponential growth). In panel (c), the dashed line shows viral kinetics in the absence of adaptive immunity (by letting $\beta_{Cn} = \beta_{Bn} = 0$; innate immunity remains active). The trajectories overlap prior to the activation of the adaptive response. The shaded region highlights the second phase (innate response). Note: changes in model parameters shifts where the three phases occur, but does not alter the underlying three-phase structure. i.e.\ existence of the three phases is robust to variation in parameters (see \emph{Supplementary Material} and Supplementary Fig.\ S3 in particular).}
\end{figure}

As reviewed by Dobrovolny \emph{et al.} \cite{Dobrovolnyetal2013}, a number of \emph{in vivo} studies have been performed to dissect the contributions of $\cd$ T cells and antibodies \cite{Iwasakietal1977,Yapetal1978,Wellsetal1981,Krisetal1988,Neffetal2003}. We use the findings of these studies to validate our model, by testing how well it is able to reproduce the experimental findings (without any further adjustment to parameters). Although the determination of the role of $\cd$ T cells is often hindered by co-inhibition of both $\cd$ T cells and the long-lived antibody response (e.g. using nude mice), it is consistently observed that antibodies play a dominant role in final viral clearance while $\cd$ T cells are primarily responsible for the timely killing of infected cells and so indirectly contribute to an increased rate of removal of free virus towards the end of infection \cite{Yapetal1978,Wellsetal1981,Fangetal2005}. Furthermore, experimental data demonstrate that a long-lived antibody response is crucial for achieving complete viral clearance, while short-lived antibodies are only capable of driving a transient decrease in viral load \cite{Iwasakietal1977,Krisetal1988}. We find that our model (with parameters calibrated against Miao \emph{et al.}'s data \cite{Miaoetal2010}) is able to reproduce these observations:
\begin{itemize}
\item virus can rebound in the absence of long-lived antibody response (see Fig.\ 5 and Supplementary Fig.\ S4).
\item both the $\cd$ T cell response and short-lived antibody response only facilitate a faster viral clearance, and are incapable of achieving clearance in the absence of long-lived antibody response (see Fig.\ 5 and Supplementary Fig.\ S4).
\item a lower level of $\cd$ T cells (modulated by a decreased level of initial naive $\cd$ T cells, $C_n$) significantly prolongs the viral clearance (see Supplementary Fig.\ S4).
\end{itemize}
In addition, the model also predicts a rapid depletion of naive $\cd$ T cells after primary infection (see Fig.\ 3), which represents a full recruitment of naive $\cd$ T cell precursors. This result may be associated with the  experimental evidence suggesting a strong correlation between the naive $\cd$ T cell precursor frequencies and effector $\cd$ T cell magnitudes for different pMHC-specific T cell populations \cite{Jenkinsetal2012}. Note that in Fig.\ 5 no adjustments to the model (e.g.\ to the vertical scale) were made; its behavior is completely determined by the calibration to the aforementioned murine data \cite{Miaoetal2010} and so these findings represent a (successful) prediction of the model.

\begin{figure}[ht!]
\centering
\includegraphics[scale=0.5]{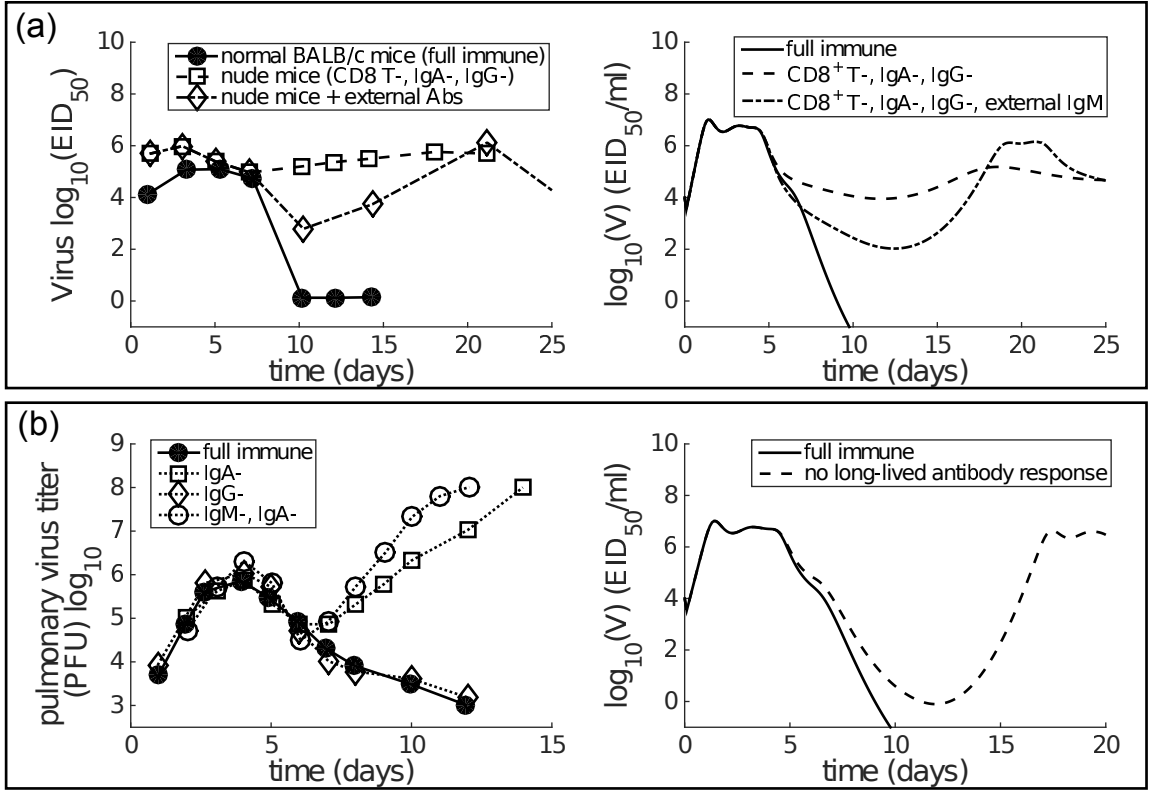}
\caption{Consistency between mice data (left panels) and model results (right panels) shows that short-lived antibody response (e.g. IgM) is only capable of generating a transient decrease in viral load while long-lived antibody response (e.g. IgA) plays a more dominant role in late-phase viral clearance. (a) Data is from the paper of Kris \emph{et al.} \cite{Krisetal1988}. Normal or nude BALB/c mice were infected with H3N2 virus. External antibodies were given at day 5 and had waned by about day 21. The model simulation mimics the passive antibody input by introducing an extra amount of IgM (in addition to antigen-stimulated IgM), whose time course faithfully follows the experimental measurement (see Fig. 2a in the paper of Kris \emph{et al.} \cite{Krisetal1988}). ``$\cd$ T-, IgA-, IgG-" was modeled by letting $\beta_{Cn} = 0$ and $p_L = 0$. ``External IgM" (in addition to the IgM produced by plasma cells) was modeled by adding a new term $-\kappa_{S} VA_e(t)$ to Eq.\ 1 where $A_e(t)$ follows a piecewise function $A_e(t) = 0$ for $t < 5$, $A_e(t) = 100(t-5)$ for $t \in [5, 7)$, $A_e(t) = 200-14(t-7)$ for $t \in [7, 21)$ and $A_e(t) = 0$ for $t \geq 21$. (b) Data is from the paper of Iwasaki \emph{et al.} \cite{Iwasakietal1977}. The data indicates that the long-lasting IgA response, but not the long-lasting IgG response or the short-lasting IgM response, is necessary for successful viral clearance. ``No long-lived antibody response" was modeled by letting $p_L = 0$. Note that Miao \emph{et al.} only measured IgM and IgG, but not IgA. As such, our model's long-lived antibody response was calibrated against IgG kinetics (see Fig.\ 2). Therefore, we emphasise that we can only investigate the relative contributions of short-lived and long-lived antibodies.}
\end{figure}

In summary, we have demonstrated that our model---with parameters calibrated against murine data \cite{Miaoetal2010}---exhibits three important phases characterized by the involvement of various immune responses. Advancing on previous models, our model does not rely on target cell depletion, and successfully reproduces a multitude of behavior from knockout experiments where particular components of the adaptive immune response were removed. This provides us with some confidence that each of the major components of the immune response has been captured adequately by our model, allowing us to now make predictions on the effect of the cellular adaptive response on viral clearance.
 
\subsection*{Dependence of the recovery time on the level of effector $\cd$ T cells}

Having established that our model is (from a structural point of view) biologically plausible and that our parameterization is capable of reproducing varied experimental data under different immune conditions (i.e.\ knockout experiments), we now study how the cellular adaptive response influences viral kinetics in detail. We focus on the key clinical outcome of recovery time, defined in the model as the time when viral titer first falls below $1\ \rm{EID_{50}/ml}$, the minimum value detected in relevant experiments (e.g. Fig.\ 2).

\begin{figure}[ht!]
\centering
\includegraphics[scale=1]{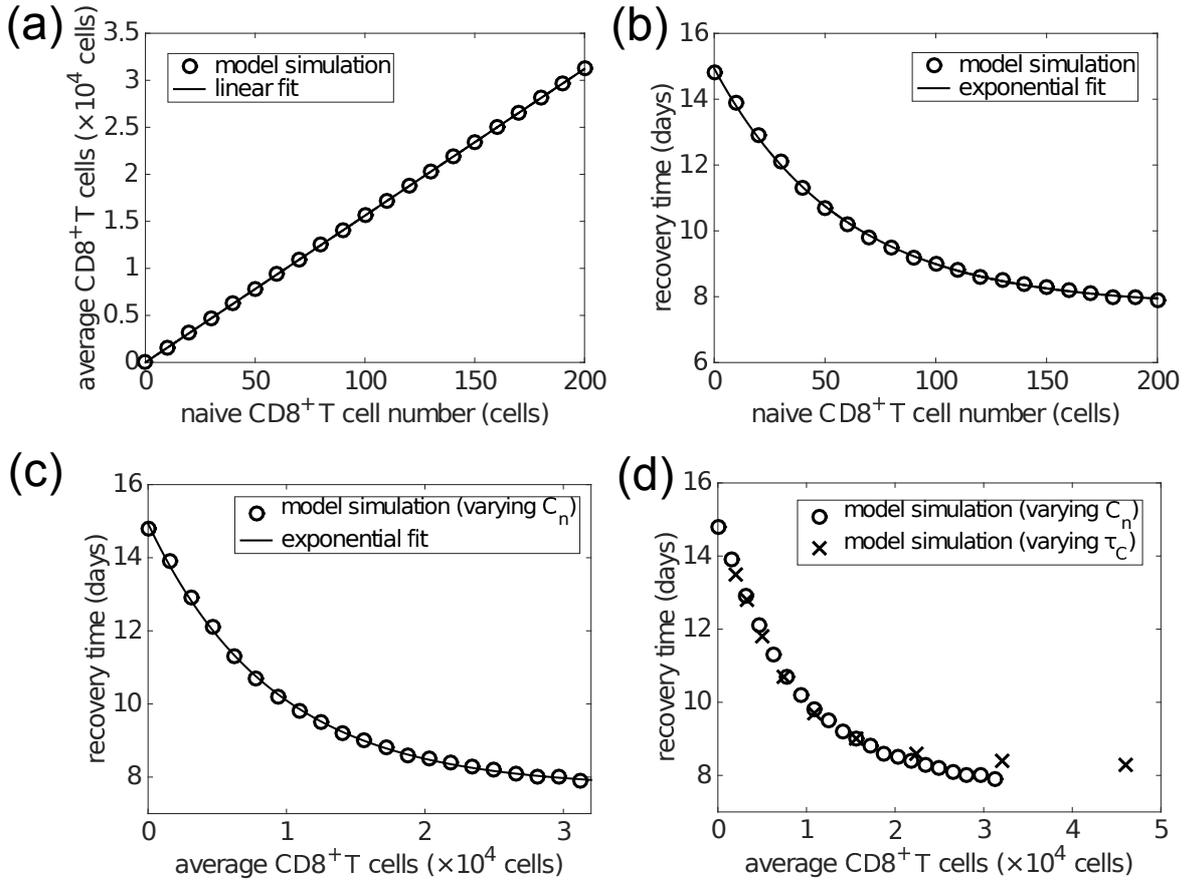}
\caption{The level of effector $\cd$ T cells plays an important role in determining recovery time. Recovery time is defined to be the time when viral load falls to 1 $\rm EID_{50}/ml$. Panel (a) shows that the average effector $\cd$ T cell number over days 6--20 is linearly related to the naive $\cd$ T cell number (i.e. $C_n(0)$). Panel (b) shows that the recovery time is approximately exponentially related to the initial naive $\cd$ T cell number. Combined, these results give panel (c) wherein an approximately exponential relationship is observed between the average $\cd$ T cell number and recovery time, both of which are experimentally measurable. Note that the exponential/linear fits shown in the figures are not generated by the viral dynamics model but are used to indicate the trends (evident visually) in the model's behavior. Panel (d) shows that varying the delay $\tau_C$ (in a similar way to that shown in Fig. S5 in the \emph{Supplementary Material}), rather than the naive $\cd$ T cell number, does not alter the exponential relationship. In panel (d), the crosses represent the results of varying $\tau_C$ and the empty circles are the same as those in panel (c) for comparison.}
\end{figure}

Time series of the viral load show that the recovery time decreases as the initial naive $\cd$ T cell number ($C_n$) increases (Supplementary Fig. S4). With that in mind, we now examine how recovery time is associated with the clinically relevant measure of effector $\cd$ T cell level during viral infection. With an increasing initial level of naive $\cd$ T cells, the average level of effector $\cd$ T cells over days 6--20 increases linearly (Fig.\ 6a), while the recovery time decreases in an approximately exponential manner (Fig.\ 6b). Combining these two effects gives rise to an approximately exponential relation between the level of effector $\cd$ T cells and recovery time (Fig.\ 6c). Note that the exponential/linear fits shown in the figures are simply to aid in interpretation of the results. They are not generated by the viral dynamics model.

If varying the delay for naive $\cd$ T cell activation and differentiation, $\tau_C$, while keeping the naive $\cd$ T cell number fixed (at the default value of 100), we find that the average level of effector $\cd$ T cells is exponentially related to the delay, while the recovery time is dependent on the delay in a piecewise linear manner (see Supplementary Fig.\ S5). Nevertheless, the combination still leads to an approximately exponential relationship between the level of effector $\cd$ T cells and recovery time (Supplementary Fig.\ S6), which is almost identical to that of varying naive $\cd$ T cells (Fig.\ 6d). We also examine the sensitivity of the exponential relationship to other model parameters generally accepted to be important in influencing the major components of the system, such as the viral production rate $p_V$, IFN production rate $p_F$ and naive B cell number. We find that the exponential relationship is robust to significant variation in all of these parameters (see Supplementary Figs.\ S6 and S7 and Fig.\ 9). These results suggest that a higher level of effector $\cd$ T cells is critical for early recovery, consistent with experimental findings \cite{Maroisetal2015}. 

\begin{figure}[ht!]
\centering
\includegraphics[scale=0.6]{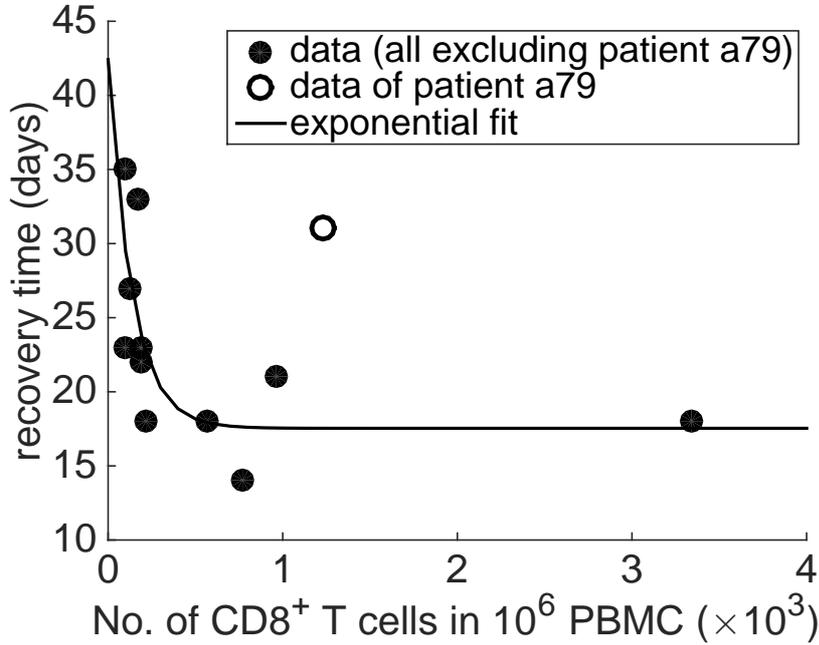}
\caption{Clinical A(H7N9) data exhibits an exponential relationship between the average $\cd$ T cell number and recovery time. The x-axis is the average level of functional effector $\cd$ T cells (i.e. $\rm IFN\gamma^+\ \cd\ T$ cells) over the period from day 8 to day 22 (or the recovery day if it comes earlier). Spearman's rank correlation test indicates a significant negative correlation between the average $\cd$ T cell numbers and recovery time ($\rho = -0.8368, p = 0.0013$). Excluding one of the patients (No. a79 in \emph{Dataset S1}; discussed in the \emph{Discussion}), all other data points (solid dots) are fitted by an offset exponential function $y = 24.8755e^{-0.0073x}+17.5356$, indicating that the best achievable recovery time for individuals with a high $\cd$ T cell response is approximately 17.5356.}
\end{figure}

Finally, and perhaps surprisingly given our model has been calibrated purely on data from the mouse, a strikingly similar relationship as shown in Fig.\ 6c is found in clinical data from influenza A(H7N9) virus-infected patients (Fig.\ 7). Excluding one patient (No. a79 in \emph{Dataset S1}; the exclusion is considered further in the \emph{Discussion}), average $\rm IFN\gamma^+\ \cd\ T$ cells and recovery time are negatively correlated (Spearman's $\rho = -0.8368$, $p = 0.0013$) and well captured by an exponential fit with an estimated offset (see Fig.\ 7 caption for details). The exponential relationship (observed in both model and data) has features of a rapid decay for relatively low/intermediate levels of effector $\cd$ T cells and a strong saturation for relatively high $\cd$ T levels, implying that even with a very high level of naive $\cd$ T cells, recovery time can not be reduced below a certain value (in this case, estimated to be approximately 17~days). Of course, the exponential relationship (i.e.\ the scale of $\cd$ T cell level or recovery time), is only a qualitative one, as we have no way to determine the scaling between different x-axis measurement units, nor adjust for particular host and/or viral factors that differ between the two experiments (i.e.\ H3N2-infection in the mouse \cite{Miaoetal2010} versus H7N9-infeciton in humans \cite{Wangetal2015}).

\subsection*{Dependence of the recovery time on the level of memory $\cd$ T cells}

In addition to naive $\cd$ T cells, memory $\cd$ T cells (established through previous viral infection) may also significantly affect recovery time due to both their rapid activation upon antigen stimulus and faster replication rate \cite{Asanoetal1996,Veigaetal2000,Badovinacetal2006,DiSpiritoetal2010}. To study the role of memory $\cd$ T cells, we must first extend our model. As we are only concerned with how the presence of memory $\cd$ T cells influences the dynamics, as opposed to the development of the memory response itself, the model is modified in a straightforward manner through addition of two additional equations which describe memory $\cd$ T cell ($C_m$) proliferation/differentiation:
\begin{align}
\frac{dC_{m}}{dt} & =-\beta_{Cm}(\frac{V}{V+h_{Cm}})C_{m}, \\
\frac{dE_m}{dt} & = \beta_{Cm}(\frac{V(t-\tau_{Cm})}{V(t-\tau_{Cm})+h_{Cm}})C_{m}(t-\tau_{Cm})e^{(p_{Cm}\tau_{Cm})} - \delta_EE_m.
\end{align}
Accordingly the term $\kappa_{E} IE$ in Eq.\ 3 is modified to $\kappa_{E} I(E+E_m)$. The full model and details on the choice of the additional parameters are provided in the \emph{Supplementary Material}. Note that the model component, $C_m$, may include different populations of memory $\cd$ T cells, including those directly specific to the virus and those stimulated by a different virus but which provide cross-protection \cite{Regner2001,Sewell2012}.

\begin{figure}[ht!]
\centering
\includegraphics[scale=1.05]{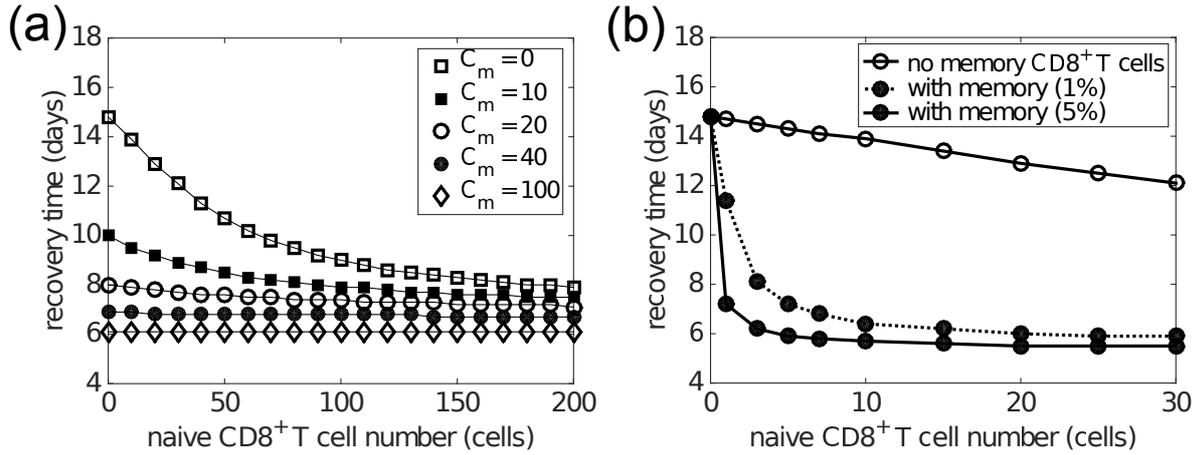}
\caption{Effects of naive and memory $\cd$ T cells on viral clearance. Recovery time is defined to be the time when the viral load falls to 1 $\rm EID_{50}/ml$. Panel (a) demonstrates that varying the number of memory $\cd$ T cells ($C_m$) reduces the recovery time for any naive $\cd$ T cell number (i.e. $C_n(0)$). Note that saturation is observed for $C_m > 100$ where the recovery time is about 6 days, independent of the naive cell numbers. Panel (b) demonstrates how the presence of pre-existing memory $\cd$ T cells (solid dots) leads to a shorter recovery time when compared to the case where no memory $\cd$ T cells are established (open circles). Note the time scale difference in panels (a) and (b). This simulation is based on the assumption that the level of pre-existing memory $\cd$ T cells is assumed to be either 1\% or 5\% (as indicated in the legend) of the maximum effector $\cd$ T cell number due to primary viral infection. The memory cell number (which is not shown in this figure) is about 30 time as many as the naive cell number shown in the figure, i.e.\ 30 naive cells result in about 900 memory cells before re-infection. }
\end{figure}
 
Fig.\ 8a shows how the pre-existing memory $\cd$ T cell number ($C_m$) changes the exponential relationship between naive $\cd$ T cells and recovery time. Importantly, as the number of memory $\cd$ T cells increases, the recovery time decreases for any level of naive $\cd$ T cells and the exponential relationship remains. The extent of reduction in the recovery time for a relatively low level of naive $\cd$ T cells is greater than that for a relatively high level of naive $\cd$ T cells. This suggests that people with a lower level of naive $\cd$ T cells may benefit more through induction of memory $\cd$ T cells, emphasizing the potential importance for taking prior population immunity into consideration when designing $\cd$ T cell-based vaccines \cite{Boltonetal2015}. 

The above result is based on the assumption that the initial memory $\cd$ T cell number upon re-infection is independent of the number of naive $\cd$ T cells available during the previous infection. However, it has also been found that the stationary level of memory $\cd$ T cells is usually maintained at about 5--10\% of the maximum antigen-specific $\cd$ T cell number during primary viral infection \cite{Muralietal1998,Hartyetal2008}. This indicates that people with a low naive $\cd$ T cell number may also develop a low level of memory $\cd$ T cells following infection. In consequence, such individuals may be relatively more susceptible to viral re-infection \cite{Houetal1994}. This alternative and arguably more realistic relationship between the numbers of naive and memory $\cd$ T cells is simulated in Fig.\ 8b where memory $\cd$ T cell levels are set to 5\% of the maximum of the effector $\cd$ T cell level. Results suggest that, upon viral re-infection, pre-existing memory $\cd$ T cells are able to significantly improve recovery time except for in hosts with a very low level of naive $\cd$ T cells (Fig.\ 8b). This is in accordance with the assumption that a smaller naive pool leads to a smaller memory pool and in turn a weaker shortening in recovery time. Although the model suggests that the failure of memory $\cd$ T cells to protect the host is unlikely to be observed (because of the approximately 30 fold increase in the size of the memory pool relative to the naive pool), the failure range may be increased if the memory pool size is much smaller (modulated by, say, changing 5\% to 1\% in the model). Therefore, for people with a low naive $\cd$ T cell number, the level of memory $\cd$ T cells may be insufficient and prior infection may provide very limited benefit, further emphasizing the opportunity for novel vaccines that are able to induce a strong memory $\cd$ T cell response to improve clinical outcomes.

\subsection*{Dependence of the recovery time on antibody level}

\begin{figure}[ht!]
\centering
\includegraphics[scale=0.6]{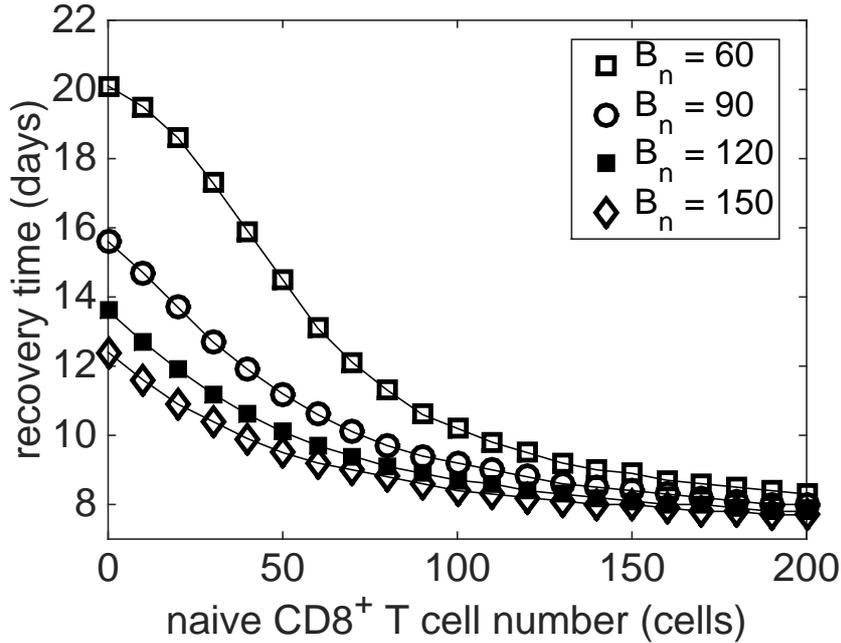}
\caption{Influence of antibody level on the relationship between the naive $\cd$ T cell number and recovery time. Recovery time is defined to be the time when viral load falls to 1 $\rm EID_{50}/ml$. Different antibody levels are simulated by varying the initial number of naive B cells (i.e. $B_n$ at $t = 0$).}
\end{figure}

Antibodies appear at a similar time as effector $\cd$ T cells during influenza viral infection and may enhance the reduction in the recovery time in addition to $\cd$ T cells. By varying the naive B cell number $B_n$ (as a convenient, but by no means unique, way to influence antibody level), we find that increasing the antibody level shortens the recovery time regardless of the initial naive $\cd$ T cell number, leaving the exponential relation largely intact (Fig.\ 9). A slight saturation occurs for the case in which levels of both naive B cells and $\cd$ T cells are low. Moreover, variation in naive B cell number also results in a wider variation in recovery time for a lower naive $\cd$ T cell level, suggesting that people with a lower level of naive $\cd$ T cells may, once again, receive a more significant benefit (in terms of recovery time) through effective induction of an antibody response via vaccination.

\section*{Discussion}

In this paper, we have studied the role of $\cd$ T cells in clearing influenza virus from the host using a viral dynamics model. The model was calibrated on a set of published murine data from Miao \emph{et al.} \cite{Miaoetal2010} and has been further shown to be able to reproduce a range of published data from experiments with different immune components knocked out. By avoiding target-cell depletion, our model is also compatible with re-infection data \cite{Laurieetal2015}, providing a strong platform on which to examine the role of $\cd$ T cells in determining recovery time from infection. Our primary finding is that the time of recovery from influenza viral infection is negatively correlated with the level of effector $\cd$ T cells in an approximately exponential manner. This robust property of infection has been identified from the model when calibrated against influenza A(H3N2) infection data in mice \cite{Miaoetal2010}, but also observed in clinical case series of influenza A(H7N9) infection in humans (Fig.\ 7) \cite{Wangetal2015}. Our findings, in conjunction with conclusions on the potential role for a T cell vaccine that stimulates and/or boosts the memory response, suggest new directions for research in both non-human species and further studies in humans on the association between $\cd$ T cell levels and clinical outcomes. Further research, including detailed statistical fitting of our model to an extensive panel of infection data (as yet unavailable) from human and non-human species, is required to establish the generality of these relationships and provide quantitative insights for specific viruses in relevant hosts.

The non-linear relationship between effector $\cd$ T cell level and recovery time may be useful in clinical treatment. The saturated property of the relation implies that a linear increase in the effector $\cd$ T cell level may result in diminishing incremental improvements in patient recovery times. With evidence of a possible age-dependent loss of naive T cells \cite{Lazuardietal2005,Cininetal2007,Cininetal2010}, our model results imply that boosting the $\cd$ T cell response via T cell vaccination may be particularly useful for those with insufficient naive $\cd$ T cells. The population-level consequences of such boosting strategies, while beyond the scope of this work, have previously been considered by the authors \cite{Boltonetal2015}.

We also investigated the effect of memory $\cd$ T cell level on viral clearance and found unsurprisingly that a high pre-existing level of memory $\cd$ T cells was always beneficial. However, our results suggest that pre-existing memory $\cd$ T cells may be particularly beneficial for certain groups of people. For example, if the memory $\cd$ T cell number induced by viral infection or vaccination is assumed to be relatively constant for everyone, people with less naive $\cd$ T cells would benefit more upon viral re-infection (see Fig.\ 8a). On the other hand, if assuming pre-existing memory $\cd$ T cell number is positively correlated with the number of naive $\cd$ T cells (simulated in Fig.\ 8b), people with more naive $\cd$ T cells would benefit more upon viral re-infection. Emerging evidence suggests that the relationship between the level of memory $\cd$ T cells and naive precursor frequencies is likely to be deeply complicated \cite{Jenkinsetal2012,LaGrutaetal2010,Thomasetal2012,Tscharkeetal2015}. In that context, our model predictions emphasize the importance for further research in this area, and the necessity to take prior population immunity into consideration when designing $\cd$ T cell vaccines \cite{Boltonetal2015}.

We modeled both short-lived and long-lived antibody responses. Experimental data and model predictions consistently show that the short-lived antibody response results in a temporary reduction in virus level whereas the long-lived antibody response is responsible for complete viral clearance (Fig.\ 5). We emphasize here that although the model is able to capture the observed short-lived and long-lived antibody responses (in order to study the virus-immune response interactions), it is not designed to investigate the mechanisms inducing different antibody responses. The observed difference in antibody decay profile may be a result of many factors including the life times of different antibody-secreting cell types \cite{Hoetal1986,Nuttetal2015}, different antibody life times \cite{Vieiraetal1988} and antibody consumption through neutralizing free virions. Detailed study of these phenomena requires a more detailed model and associated data for parameter estimation and model validation, and is thus left for future work. Similarly, $\rm CD4^+$ T cells are also known to perform a variety of functions in the development of immunity, such as facilitation of the formation and optimal recall of $\cd$ T cells or even direct killing of infected cells during viral infection \cite{Wherryetal2004,LaGrutaetal2014,Riberdyetal2000,Laidlawetal2014}. Their depletion due to, say, HIV infection has also been associated with more severe clinical outcomes following influenza infection \cite{Cohenetal2015}. Some of the major functions of $\rm CD4^+$ T cells may be considered to be implicitly modeled through relevant parameters such as the rate of recall of memory $\cd$ T cells (modeled by the delay $\tau_{Cm}$) in our extended model which includes memory $\cd$ T cells. However, a detailed viral dynamics study of the role of $\rm CD4^+$ T cells in influenza infection, including in HIV infected patients with depleted $\rm CD4^+$ T cells, remains an open and important challenge.

In a recent theoretical study, it was found that spatial heterogeneity in the T cell distribution may influence viral clearance \cite{Zarnitsynaetal2016}. Resident $\cd$ T cells in the lungs have a more direct and significant effect on timely viral clearance than do naive and memory pools resident in lymph nodes. Although this factor has been partially taken into consideration in our model by introducing a delay for naive/memory $\cd$ T cells, lack of explicit modeling of the spatial dynamics limits a direct application of our model to investigate these spatial effects. 

Finally, as noted in the results, one of the influenza A(H7N9)-infected patients (patient a79) was not included in our analysis of the clinical data (Fig.\ 7). Although our model suggests some possibilities for the source of variation due to possible variation in parameter values, large variations in recovery time are only expected to occur for relatively low levels of naive $\cd$ T cells, nominally incompatible with this patient's moderate $\cd$ T cell response but a relatively long recovery time. However, we note that $\rm IFN\gamma^+\ \cd$ T cell counts for this patient were only collected at days 10 and 23, and that the count at day 10 was particularly low and much lower than that at day 23 (see \emph{Dataset S1}). We suspect that delayed, rather than weakened, production (to at least day 10) of the $\rm IFN\gamma^+\ \cd$ T cell response in this patient substantially contributed to the observed delay in recovery. Further investigation of this patient's clinical course and clinical samples is currently being undertaken.

\section*{Funding}

PC is supported by an Australian Government National Health and Medical Research Council (NHMRC) funded Centre for Research Excellence in Infectious Diseases Modelling to Inform Public Health Policy. ZW is supported by an NHMRC Australia--China Exchange Fellowship. AWCY is supported by an Australian Postgraduate Award. JM and KK are support by NHMRC Career Development and Senior Researcher Fellowships respectively. JMM is supported by an Australian Research Council Future Fellowship. The H7N9 clinical study was supported by National Natural Science Foundation of China (NFSC) grants 81471556, 81470094 and 81430030.

\section*{Acknowledgements}

The authors would like to thank members of the Modelling and Simulation Unit in the Centre for Epidemiology and Biostatistics and the School of Mathematics and Statistics, University of Melbourne, for helpful advice on the study.

%

\section*{Competing interests}

We have no competing interests.

\newpage
\begin{table}[!ht]
\begin{center}
\begin{tabular}{p{0.6cm}p{8.9cm}p{3.3cm}p{1.6cm}}
   \hline
   {\bf Par.} & {\bf Description} & {\bf Value} \& {\bf Unit} & {\bf Ref.}\\
   \hline
   $V_0$ & initial viral load & $10^{4}$ $\rm [u_V]$ & estimated \\
   \hline
   $T_0$ & initial number of epithelial cells in the URT & $7\times10^{7}$ cells & \cite{Petrieetal2013} \\
   \hline
   $g_T$ & base growth rate of healthy cells & $0.8\ \rm d^{-1}$ & fixed \\
   \hline
   $p_V$ & viral production rate & $210\ \rm [u_V]cell^{-1}d^{-1}$ & estimated \\
 \hline
   $p_F$ & IFN production rate & $10^{-5}\ \rm [u_F]cell^{-1}d^{-1}$ & estimated \\
 \hline
   $p_C$ & naive $\cd$ T cell proliferation rate & $1.2\ \rm d^{-1}$ &  \cite{Bocharovetal1994}\\
   \hline
   $p_B$ & naive B cell proliferation rate & $0.52\ \rm d^{-1}$ & estimated \\
   \hline
   $p_S$ & short-lived antibody production rate & $12\ \rm [u_A]cell^{-1}d^{-1}$ & estimated \\
   \hline
   $p_L$ & long-lived antibody production rate & $4\ \rm [u_A]cell^{-1}d^{-1}$ & estimated \\
   \hline
   $\delta_V$ & nonspecific viral clearance rate & 5 $\rm d^{-1}$ & \cite{Baccametal2006}\\
   \hline
   $\delta_I$ & nonspecific death rate of infected cells & 2 $\rm d^{-1}$ & \cite{Bocharovetal1994} \\
   \hline
   $\delta_F$ & IFN degradation rate & 2 $\rm d^{-1}$ & \cite{Paweleketal2012} \\
   \hline
   $\delta_E$ & death rate of effector $\cd$ T cells & 0.57 $\rm d^{-1}$ & \cite{Veigaetal2000} \\
   \hline
   $\delta_P$ & death rate of plasma B cells & 0.5 $\rm d^{-1}$ & estimated \\
   \hline
   $\delta_S$ & short-lived antibody consumption rate & 2 $\rm d^{-1}$ & estimated \\
   \hline
   $\delta_L$ & long-lived antibody consumption rate & 0.015 $\rm d^{-1}$ & estimated \\
   \hline
   $\beta$ & rate of viral consumption by binding to target cells & $5\times10^{-7}\ \rm cell^{-1}d^{-1}$ & estimated \\
   \hline
   $\beta'$ & rate of infection of target cells by virus & $3\times10^{-8}\ \rm [u_V]^{-1}d^{-1}$ & estimated \\
   \hline
   $\phi$ & rate of conversion to virus-resistant state & 0.33 $\rm [u_F]^{-1}d^{-1}$ & \cite{Paweleketal2012} \\
   \hline
   $\rho$ & rate of recovery from virus-resistant state & 2.6 $\rm d^{-1}$ & \cite{Paweleketal2012} \\
   \hline
   $\kappa_S$ & rate of viral neutralization by short-lived antibodies & 0.8 $\rm [u_A]^{-1}d^{-1}$ & estimated \\
   \hline
   $\kappa_L$ & rate of viral neutralization by long-lived antibodies & 0.4 $\rm [u_A]^{-1}d^{-1}$ & estimated \\
   \hline
   $\kappa_N$ & killing rate of infected cells by IFN-activated NK cells & 2.5 $\rm [u_F]^{-1}d^{-1}$ & \cite{Paweleketal2012}\\
   \hline
   $\kappa_E$ & killing rate of infected cells by effector $\cd$ T cells & $5\times10^{-5}\ \rm cell^{-1}d^{-1}$ & estimated\\
   \hline
   $\beta_{Cn}$ & maximum rate of stimulation of naive $\cd$ T cells by virus & 1 $\rm d^{-1}$ & \cite{Wherryetal2004} \\
   \hline
   $\beta_{Bn}$ & maximum rate of stimulation of naive B cells by virus & 0.03 $\rm d^{-1}$ &  estimated\\
   \hline
   $h_C$ & half-maximal stimulating viral load for naive $\cd$ T cells & $10^4\ \rm [u_V]$ & estimated  \\
   \hline
   $h_B$ & half-maximal stimulating viral load for naive B cells & $10^4\ \rm [u_V]$ & estimated \\
   \hline
   $\tau_C$ & delay for effector $\cd$ T cell production & 6 d & \cite{Lawrenceetal2005}\\
   \hline
   $\tau_B$ & delay for plasma B cell production & 4 d & estimated \\
   \hline
\end{tabular}
\caption{Model parameter values obtained by fitting the model to experimental data. $[u_V]$, $[u_F]$ and $[u_A]$ represent the units of viral load, IFN and antibodies respectively. $[u_V]$ and $[u_A]$ are $\rm EID_{50}/ml$ (50\% egg infective dose) and $\rm pg/ml$, consistent with the units of data. IFN is assumed to be a non-dimensionalized variable in the model, and therefore $[u_F]$ can be ignored. Some parameters are obtained from the literature and the rest are obtained by fitting the model to experimental data in the paper of Miao \emph{et al.} \cite{Miaoetal2010}, except $g_T$ which is of minor importance when considering a single infection and is thus fixed to reduce uncertainty.}
\end{center}
\end{table}

\newpage
\section*{Supplementary Material}

\subsection{Numerical method for solving the model}

The delay in activation of $\cd$ T cells and B cells directly results in the delay of production of effector $\cd$ T cells and antibody-producing cells. By a time-shift transform $t = t'+\tau_C$, Eq.\ 7 in the main text becomes
\begin{equation}
\frac{dE(t'+\tau_C)}{dt'}  = \beta_{Cn}(\frac{V(t')}{V(t')+h_C})C_{n}(t')e^{(p_C\tau_C)} - \delta_EE(t'+\tau_C), \tag{S1}
\end{equation}
where $t \geq 0$ and $t'$ starts from $-\tau_C$. As defined in the main text, $V = 0$ for any negative time, i.e. $V(t') = 0$ for $-\tau_C< t' < 0$. Moreover, $E(t'+\tau_C) = 0$ for $\tau_C< t' < 0$. Thus, the equation is trivial for $t' \geq 0$. Therefore, for $t' \geq 0$, we replace $t'$ back to $t$ and Eq.\ S1 becomes
\begin{equation}
\frac{dE(t+\tau_C)}{dt}  = \beta_{Cn}(\frac{V(t)}{V(t)+h_C})C_{n}(t)e^{(p_C\tau_C)} - \delta_EE(t+\tau_C), \tag{S2}
\end{equation}
where $t \geq 0$. Similarly, Eqs.\ 9-11 in the main text can be changed to
\begin{align}
\frac{dP(t+\tau_B)}{dt} &  = \beta_{Bn}(\frac{V(t)}{V(t)+h_B})B_{n}(t)e^{(p_B\tau_B)} - \delta_PP(t+\tau_B), \tag{S3}\\
\frac{dA_S(t+\tau_B)}{dt} & =p_SP(t+\tau_B)-\delta_SA_S(t+\tau_B), \tag{S4} \\
\frac{dA_L(t+\tau_B)}{dt} & =p_LP(t+\tau_B)-\delta_LA_L(t+\tau_B), \tag{S5}
\end{align}
In this way, the model presented in the main text is equivalent to the following model:
\begin{align}
\frac{dV}{dt} & =p_VI-\delta_VV-\kappa_{S} VA_S(t)-\kappa_{L} VA_L(t)-\beta VT,   \tag{S6}\\
\frac{dT}{dt} & =g_T(T+R)(1-\frac{T+R+I}{T_0})-\beta' VT+\rho R-\phi FT,  \tag{S7}\\
\frac{dI}{dt} & =\beta' VT-\delta_I I-\kappa_{N} IF-\kappa_{E} IE(t),  \tag{S8}\\
\frac{dF}{dt} & =p_FI-\delta_FF,  \tag{S9}\\
\frac{dR}{dt} & =\phi FT-\rho R,  \tag{S10}\\
\frac{dC_{n}}{dt} & =-\beta_{Cn}(\frac{V}{V+h_C})C_{n},  \tag{S11}\\
\frac{dE(t+\tau_C)}{dt} & = \beta_{Cn}(\frac{V}{V+h_C})C_{n}e^{(p_C\tau_C)} - \delta_EE(t+\tau_C), \tag{S12}\\
\frac{dB_n}{dt} & =-\beta_{Bn}(\frac{V}{V+h_B})B_{n}, \tag{S13} \\
\frac{dP(t+\tau_B)}{dt} & = \beta_{Bn}(\frac{V}{V+h_B})B_{n}e^{(p_B\tau_B)} - \delta_PP(t+\tau_B), \tag{S14}\\
\frac{dA_S(t+\tau_B)}{dt} & =p_SP(t+\tau_B)-\delta_SA_S(t+\tau_B), \tag{S15} \\
\frac{dA_L(t+\tau_B)}{dt} & =p_LP(t+\tau_B)-\delta_LA_L(t+\tau_B). \tag{S16}
\end{align}
For variables whose independent variable is not explicitly specified, they are all functions of $t$, i.e. $V$ reads $V(t)$. This model avoids negative time and is solved by the following steps:
\begin{itemize}
\item Firstly choosing a time step size $\Delta t$ and using it to discretise the time domain to be $t = 0, \Delta t, 2\Delta t, 3\Delta t, ... , N\Delta t$. The results in the main text are generated using $\Delta t = 0.1$ (day), the choice of which is based on the result that further decreasing $\Delta t$ does not improve the solution (results not shown). 
\item Given initial condition 

$(V,T,I,F,R,C_n,E,B_n,P,A_S,A_L) = (V_0,7\times10^7,0,0,0,100,0,100,0,0,0)$, 

we solve the model iteratively. For iteration from $k\Delta t$ to $(k+1)\Delta t$, $A_S(t)$, $A_L(t)$ and $E(t)$ in Eqs.\ S6 and S8 are already known and thus treated as parameters. The system becomes an ODE system and can be easily solved by using a built-in ODE solver \emph{ode15s} in MATLAB(R2014b) with default settings. All the variables at time $(k+1)\Delta t$ are then updated for use in the next iteration.  

\end{itemize} 
MATLAB code is provided below.

\begin{lstlisting}[frame=single]
clear
dt=0.1; % time step size
time=0:dt:100;
 
% parameters
T0=7e+7;gT=0.8;pV=210;deltaV=5.0;
beta=5e-7;betap=3e-8;
deltaI=2;kappaN=2.5;kappaE=5e-5;
phi=0.33;rho=2.6;pF=1e-5;deltaF=2;
betaCn=1;betaBn=0.03;
kappaS=0.8;kappaL=0.4;
pC=1.2;pB=0.52;deltaE=0.57;deltaP=0.5;
pS=12;pL=4;deltaS=2;deltaL=0.015; 
tauC=6;tauB=4;hC=1e+4;hB=1e+4;
 
% index indicating when delayed process starts
indC=round(tauC/dt+1); % for tauC
indB=round(tauB/dt+1); % for tauB
 
% variable vectors and initial conditions
V=zeros(1,length(time));V(1)=1e+4;
T=V;T(1)=7e+7;
I=T;I(1)=0;
R=I;
F=I;
Cn=100*ones(1,length(time));
Bn=100*ones(1,length(time));
E=zeros(1,indC+length(time));
P=I;
AS=zeros(1,indB+length(time));
AL=AS;
 
init=[V(1),T(1),I(1),R(1),F(1),Cn(1),E(1),Bn(1),P(1),AS(1),AL(1)]';
 
options = odeset('RelTol',1e-3,'AbsTol',1e-6);
 
for i=2:length(time) 
    [~,Y]=ode15s(@ODEmodel,[0 dt],init,options,E(i),AL(i),AS(i),tauC,tauB,phi,rho,deltaF,gT,pF,pV,beta,betap,kappaN,deltaV,deltaI,betaCn,betaBn,kappaE,kappaS,pL,pS,deltaL,deltaS,deltaP,deltaE,pC,pB,kappaL,hC,hB,T0);
    
    V(i)=Y(end,1);T(i)=Y(end,2);I(i)=Y(end,3);R(i)=Y(end,4);
    F(i)=Y(end,5);Cn(i)=Y(end,6);Bn(i)=Y(end,8);P(i)=Y(end,9);
    E(indC+i)=Y(end,7);
    AS(indB+i)=Y(end,10);
    AL(indB+i)=Y(end,11);
 
    init=Y(end,:)'; % initial condition for next iteration
end

\end{lstlisting}

In the command of \emph{ode15s}, a function ``ODEmodel" is required and provided below.

\begin{lstlisting}[frame=single]
function ynew=ODEmodel(~,y,E,AL,AS,tauC,tauB,phi,rho,deltaF,gT,pF,pV,beta,betap,kappaN,deltaV,deltaI,betaCn,betaBn,kappaE,kappaS,pL,pS,deltaL,deltaS,deltaP,deltaE,pC,pB,kappaL,hC,hB,T0)
 
% V: viral load
% T: target cell
% I: infected cell
% R: Resistant cell
% F: IFN
% Cn: naive CD8+ T cells
% E:  effector CD8+ T cells
% Bn: naive B cells
% P:  plasma B cells
% AS: short-lived antibodies
% AL: long-lived antibodies
 
% y=[V,T,I,R,F,Cn,E,Bn,P,AS,AL]
 
ynew=zeros(11,1);
 
ynew(1)=pV*y(3)-deltaV*y(1)-kappaS*y(1)*AS-kappaL*y(1)*AL-beta*y(1)*y(2);
ynew(2)=gT*(y(2)+y(4))*(1-(y(2)+y(3)+y(4))/T0)-betap*y(1)*y(2)+rho*y(4)-phi*y(2)*y(5);
ynew(3)=betap*y(1)*y(2)-deltaI*y(3)-kappaN*y(3)*y(5)-kappaE*y(3)*E;
ynew(4)=phi*y(2)*y(5)-rho*y(4);
ynew(5)=pF*y(3)-deltaF*y(5);
ynew(6)=-betaCn*y(1)/(y(1)+hC)*y(6);
ynew(7)=betaCn*y(1)/(y(1)+hC)*y(6)*exp(pC*tauC)-deltaE*y(7);
ynew(8)=-betaBn*y(1)/(y(1)+hB)*y(8);
ynew(9)=betaBn*y(1)/(y(1)+hB)*y(8)*exp(pB*tauB)-deltaP*y(9);
ynew(10)=pS*y(9)-deltaS*y(10);
ynew(11)=pL*y(9)-deltaL*y(11);
\end{lstlisting}

\newpage

\subsection{Details of fitting the model to data}

The model contains 11 equations and 30 parameters (see Table 1 in the main text). This represents a serious challenge in terms of parameter estimation, and clearly prevents a straightforward application of standard statistical techniques. However, based on an extensive survey of the experimental literature, we have been able to identify plausible, but by no means unique, combinations of parameters that successfully explain the available data. A number of parameters were taken directly from the literature, as per the citations in Table\ 1. The rest (18 parameters) were estimated by calibrating the model to the published data from Miao \emph{et al.} \cite{Miaoetal2010} who measured viral titre, $\cd$ T cell counts and IgM and IgG antibodies in laboratory mice (exhibiting a full immune response) over time during primary influenza H3N2 virus infection (shown in Fig.\ 2 in the main text). Note that the data were presented in scatter plots in the original paper \cite{Miaoetal2010}, while we presented the data in Mean $\pm$ SD at each data collection time point (as shown in Fig.\ 2 in the main text) and fit our mean-field mathematical model to the means. 

To obtain the set of parameters used for the main analysis from the experimental data in the paper of Miao \emph{et al.} \cite{Miaoetal2010}, we took the following approach:
\begin{itemize}
\item We first manually determined a set of the 18 parameters which produced a model solution that reasonably matched the experimental data shown in Fig.\ 2 in the main text. In detail, the main criteria include: 1) the viral load starts from about $10^4 \rm EID_{50}/ml$, reaches a peak of about $10^7 \rm EID_{50}/ml$ at 1--2 days p.i., then declines rapidly from about 5 days p.i. (note that the last three data points were not considered due to the limit of detection); 2) the $\cd$ T cell count starts to increase rapidly at about 6 days p.i., reaches a peak of about $10^5$--$10^6$ at 8--10 days p.i. and returns back to zero after about 20 days p.i.; 3) the IgM level starts to increase rapidly at 4--5 days p.i., reaches a peak of about 200--300 pg/ml at about 10 days p.i. and returns back to baseline after about 20 days p.i.; 4) the IgG level starts to increase rapidly at 4--5 days p.i., reaches a peak of about 800--1000 pg/ml at about 20 days p.i. and decays slowly. Given the high-dimensionality of the parameter space and limited experimental data, the procedure was essential in allowing us to identify a candidate parameter set which was not far from generating a local minimum in the following optimisation process.

\item The candidate parameter set was then used as an initial estimate for optimization using MATLAB's built-in function \emph{fmincon} with default settings. The target of optimization was to minimize the least-squares error (LSE):
\begin{equation}
L = L_V + L_E +L_{M} +L_{G}, \tag{S17}
\end{equation}
where the four LSE components for viral load, effector $\cd$ T cells, IgM and IgG were given respectively by
\begin{equation}
L_V = \sum_i [(V_{model}(t_i) - V_{data}(t_i))w_V(t_i)/s_V]^2, \tag{S18}
\end{equation}
\begin{equation}
L_E = \sum_i [(E_{model}(t_i) - E_{data}(t_i))w_E(t_i)/s_E]^2, \tag{S19}
\end{equation}
\begin{equation}
L_M = \sum_i [(M_{model}(t_i) - M_{data}(t_i))w_M(t_i)/s_M]^2, \tag{S20}
\end{equation}
\begin{equation}
L_G = \sum_i [(G_{model}(t_i) - G_{data}(t_i))w_G(t_i)/s_G]^2. \tag{S21}
\end{equation}
$V_{model}$, $E_{model}$, $M_{model}$ and $G_{model}$ indicate the model solution for variables $V$, $E$, $A_S$ and $A_L$ evaluated at time $t_i$ respectively. $V_{data}$, $E_{data}$, $M_{data}$ and $G_{data}$ indicate the associated data. $i$ is the index of the time point and the $t_i$ may differ for the four components. $s_V = 10^7$,  $s_E = 6\times 10^4$,  $s_M = 300$ and  $s_G = 900$ were used to scale the errors to the same order of magnitude. Due to the fact that data were not collected at a fixed frequency (i.e. the time interval between adjacent data points is not constant), the errors at different time points were assigned different weights based on the length of time intervals between adjacent points. For example, if viral load was measured at time $t_i, (i = 0,1,2,...,k)$, the weight function $w_V(t_i)$ for interior points was given by
\begin{equation}
w_V(t_i)= \frac{t_{i+1}-t_{i-1}}{2(t_{k}-t_0)}, \tag{S22}
\end{equation}
and for boundary points by 
\begin{equation}
w_V(t_0)= \frac{t_{1}-t_{0}}{2(t_{k}-t_0)} \qquad  {\rm and} \qquad  w_V(t_k)= \frac{t_{k}-t_{k-1}}{2(t_{k}-t_0)}. \tag{S23}
\end{equation}
This error weighting was used to weaken the domination of dense data points on model fits. It is evident that a lot of measurements were done within the first 10 days post-infection but only a few were performed after day 20 post-infection, in particular for IgG data (see Fig.\ 2 in the main text). We found that using equally weighted LSEs led to a model fit that manifestly failed to capture those sparse data points which we believe are equally, if not more important from a more biological perspective, in determining the IgG kinetics (see Fig.\ S8, compared with Fig.\ 2 in the main text).  

\item The parameter constraints when using \emph{fmincon} were set to $V_0 \in [10^3, 10^5]$, $p_V \in [50, 500]$, $\beta \in [0, 1]$, $\beta' \in [0, 1]$, $p_F \in [0, 1]$, $\kappa_S \in [0, 2]$, $\kappa_L \in [0, 1]$, $\kappa_E \in [0, 1]$, $\beta_{Bn} \in [0, 2]$, $p_B \in [0, 1]$, $\delta_P \in [0, 1]$, $\delta_S \in [0, 5]$, $p_S \in [0, 20]$, $p_L \in [0, 20]$, $\delta_L \in [0, 0.1]$, $\tau_B \in [3, 7]$, $h_C \in [10^3, 10^5]$ and $h_B \in [10^3, 10^5]$.
 
\item After obtaining a locally optimized solution for the candidate parameter set, we then checked the solution generated and evaluated its biological plausibility (based on the criteria mentioned above). This step was essential as given the over-specification of the model (in a statistical sense), it was possible for good fitting solutions to be identified by MATLAB's optimization algorithm, which were nonetheless biologically implausible. For example, oscillatory solutions for quantities such as IgG, while providing a ``good-fit" to data, were not deemed acceptable on biological grounds (see Fig.\ S9 for such an example). If the optimised solution failed our (qualitative) evaluation, we returned to the first step and redetermined a new set of parameters as new initial estimates to be optimized using MATLAB.
\end{itemize}
This entire process was repeated to arrive at the default parameter set shown in Table\ 1 in the main text. To guarantee that the default parameter set was a good choice, we further randomly generated 10,000 sets of parameter samples near the default parameter set (within $\pm$50\% from the default values) and used them as initial estimates with \emph{fmincon} to search for locally optimized solutions. Of these, 31  generated a better LSE but all failed to meet the criteria mentioned above (results not shown).   

Fig.\ 2 in the main text shows how the model reproduces the key dynamic behavior shown in the data. We also show in the \emph{Results} section in the main text, that the model behavior is robust to perturbation of model parameters and that model predictions are reasonably consistent with a range of other experimental data, demonstrating the plausibility, if not uniqueness (of course), of the parameter set. We emphasize that, although the default parameter set is successful in reproducing a multitude of experimental observations (e.g. full immune, knockout, re-infection) as presented in the main text, we by no means claim that this parameter set is unique. It remains an open and challenging problem to reliably identify a biologically plausible and statistically identifiable solution for what is a highly complex system, where we are severely limited by available experimental data.

\newpage
\subsection{The model with memory $\cd$ T cells}

Incorporating memory $\cd$ T cells into the model in the main text, we only make two changes. The first is adding two equations to describe the memory $\cd$ T cell ($C_m$) proliferation/differentiation, similar to Eqs.\ 6 and 7 in the main text
\begin{align}
\frac{dC_{m}}{dt} & =-\beta_{Cm}(\frac{V}{V+h_{Cm}})C_{m}, \tag{S24}\\
\frac{dE_m}{dt} & = \beta_{Cm}(\frac{V(t-\tau_{Cm})}{V(t-\tau_{Cm})+h_{Cm}})C_{m}(t-\tau_{Cm})e^{(p_{Cm}\tau_{Cm})} - \delta_EE_m. \tag{S25}
\end{align}
Then we change the term $\kappa_{E} IE$ in Eq.\ 3 in the main text to $\kappa_{E} I(E+E_m)$. Hence, similar to the approach mentioned above that moving the delayed term from viral load to effector cells, we write down the model in an equivalent form,  
\begin{align}
\frac{dV}{dt} & =p_VI-\delta_VV-\kappa_{S} VA_S(t)-\kappa_{L} VA_L(t)-\beta VT,   \tag{S26}\\
\frac{dT}{dt} & =g_T(T+R)(1-\frac{T+R+I}{T_0})-\beta' VT+\rho R-\phi FT,  \tag{S27}\\
\frac{dI}{dt} & =\beta' VT-\delta_I I-\kappa_{N} IF-\kappa_{E} I[E(t)+E_m(t)],  \tag{S28}\\
\frac{dF}{dt} & =p_FI-\delta_FF,  \tag{S29}\\
\frac{dR}{dt} & =\phi FT-\rho R,  \tag{S30}\\
\frac{dC_{n}}{dt} & =-\beta_{Cn}(\frac{V}{V+h_C})C_{n},  \tag{S31}\\
\frac{dE(t+\tau_C)}{dt} & = \beta_{Cn}(\frac{V}{V+h_C})C_{n}e^{(p_C\tau_C)} - \delta_EE(t+\tau_C), \tag{S32}\\
\frac{dB_n}{dt} & =-\beta_{Bn}(\frac{V}{V+h_B})B_{n}, \tag{S33} \\
\frac{dP(t+\tau_B)}{dt} & = \beta_{Bn}(\frac{V}{V+h_B})B_{n}e^{(p_B\tau_B)} - \delta_PP(t+\tau_B), \tag{S34}\\
\frac{dA_S(t+\tau_B)}{dt} & =p_SP(t+\tau_B)-\delta_SA_S(t+\tau_B), \tag{S35} \\
\frac{dA_L(t+\tau_B)}{dt} & =p_LP(t+\tau_B)-\delta_LA_L(t+\tau_B). \tag{S36}\\
\frac{dC_{m}}{dt} & =-\beta_{Cm}(\frac{V}{V+h_{Cm}})C_{m},  \tag{S37}\\
\frac{dE_m(t+\tau_{Cm})}{dt} & = \beta_{Cm}(\frac{V}{V+h_{Cm}})C_{m}e^{(p_{Cm}\tau_{Cm})} - \delta_EE_m(t+\tau_{Cm}). \tag{S38}
\end{align}

For variables whose independent variables are not explicitly specified, they are all functions of $t$, i.e. $V$ reads $V(t)$. Memory $\cd$ T cells show a shorter delay and faster proliferation than naive $\cd$ T cells \cite{Veigaetal2000}. The shortened delay may be caused by a shortened lag time to the first division and/or a reduced delay for effector cells migrating from the lymphatic compartment to the lung \cite{Veigaetal2000,Leeetal2009,Zarnitsynaetal2016}. The former reduction is about 15 hours \cite{Veigaetal2000} and the latter is less than about 12 hours \cite{Leeetal2009}. Thus, we choose $\tau_{Cm} = 5$ (days), correspond to a one day reduction in the delay time compared to the delay of naive cells ($\tau_C = 6$). Memory $\cd$ T cells show a higher division rate and a lower loss rate than naive $\cd$ T cells \cite{Veigaetal2000}, based on which the net production rate of effector cells for memory $\cd$ T cells is estimated to be about 1.5 times of that for naive cells. Thus,we choose $p_{Cm} = 1.5p_{C} = 1.8$ ($\rm day^{-1}$). In the absence of data, we assume $\beta_{Cm} = \beta_{C}$ and $h_{Cm} = h_C$. The initial number of memory $\cd$ T cells is varied as specified in the main text or figures. We assume that the effector $\cd$ T cells produced by either naive or memory $\cd$ T cells are functionally identical (i.e. then have the same decay rate $\delta_E$ and killing rate $\kappa_E$). Note that we do not model the process of differentiation of effector $\cd$ T cells into memory cells but use a memory cell pool as an initial condition to simulate viral re-infection.

MATLAB code is provided below.

\begin{lstlisting}[frame=single]
clear
dt=0.1; % time step size
time=0:dt:100;
% parameters
T0=7e+7;gT=0.8;pV=210;deltaV=5.0;
beta=5e-7;betap=3e-8;
deltaI=2;kappaN=2.5;kappaE=5e-5;
phi=0.33;rho=2.6;pF=1e-5;deltaF=2;
betaCn=1;betaBn=0.03;
kappaS=0.8;kappaL=0.4;
pC=1.2;pB=0.52;deltaE=0.57;deltaP=0.5;
pS=12;pL=4;deltaS=2;deltaL=0.015; 
tauC=6;tauB=4;hC=1e+4;hB=1e+4;
betaCm=1;pCm=1.8;tauCm=5;
 
% index indicating when delayed process starts
indC=round(tauC/dt+1); % for tauC
indCm=round(tauCm/dt+1); % for tauCm
indB=round(tauB/dt+1); % for tauB
 
% variable vectors and initial conditions
V=zeros(1,length(time));V(1)=1e+1;
T=V;T(1)=7e+7;I=T;I(1)=0;R=I;F=I;
Cn=100*ones(1,length(time));
Bn=100*ones(1,length(time));
E=zeros(1,indC+length(time));
P=I;AS=zeros(1,indB+length(time));AL=AS; 
Cm=5000*ones(1,length(time));
Em=zeros(1,indCm+length(time));
 
init=[V(1),T(1),I(1),R(1),F(1),Cn(1),E(1),Bn(1),P(1),AS(1),AL(1),
     Cm(1),Em(1)]';
 
options = odeset('RelTol',1e-3,'AbsTol',1e-6);
 
for i=2:length(time) 
    [~,Y] = ode15s(@ODEmodel_with_memory,[0 dt],init,options,E(i),AL(i),AS(i),Em(i),tauC,tauB,phi,rho,deltaF,gT,pF,pV,beta,betap,kappaN,deltaV,deltaI,betaCn,betaBn,kappaE,kappaS,pL,pS,deltaL,deltaS,deltaP,deltaE,pC,pB,kappaL,hC,hB,T0,betaCm,pCm,tauCm);
    
    V(i)=Y(end,1);T(i)=Y(end,2);I(i)=Y(end,3);R(i)=Y(end,4);
    F(i)=Y(end,5);Cn(i)=Y(end,6);Bn(i)=Y(end,8);P(i)=Y(end,9);
    E(indC+i)=Y(end,7);
    AS(indB+i)=Y(end,10);AL(indB+i)=Y(end,11);
    Cm(i)=Y(end,12);Em(indCm+i)=Y(end,13);
    init=Y(end,:)'; % initial condition for next iteration
end
\end{lstlisting}
The function ``ODEmodel\_with\_memory" is provided below.

\begin{lstlisting}[frame=single]
function ynew=ODEmodel_New_with_memory(~,y,E,AL,AS,Em,tauC,tauB,phi,rho,deltaF,gT,pF,pV,beta,betap,kappaN,deltaV,deltaI,betaCn,betaBn,kappaE,kappaS,pL,pS,deltaL,deltaS,deltaP,deltaE,pC,pB,kappaL,hC,hB,T0,betaCm,pCm,tauCm)
 
% V: viral load
% T: target cell
% I: infected cell
% R: Resistant cell
% F: IFN
% Cn: naive CD8+ T cells
% E:  effector CD8+ T cells
% Bn: naive B cells
% P:  plasma B cells
% AS: short-lived antibodies
% AL: long-lived antibodies
% Cm:  memory CD8+ T cells
% Em:  effector CD8+ T cells produced from memory cells
 
% y=[V,T,I,R,F,Cn,E,Bn,P,AS,AL,Cm,Em]
 
ynew=zeros(13,1);
ynew(1)=pV*y(3)-deltaV*y(1)-kappaS*y(1)*AS-kappaL*y(1)*AL-beta*y(1)*y(2);
ynew(2)=gT*(y(2)+y(4))*(1-(y(2)+y(3)+y(4))/T0)-betap*y(1)*y(2)+rho*y(4)-phi*y(2)*y(5);
ynew(3)=betap*y(1)*y(2)-deltaI*y(3)-kappaN*y(3)*y(5)-kappaE*y(3)*(E+Em);
ynew(4)=phi*y(2)*y(5)-rho*y(4);
ynew(5)=pF*y(3)-deltaF*y(5);
ynew(6)=-betaCn*y(1)./(y(1)+hC)*y(6);
ynew(7)=betaCn*y(1)./(y(1)+hC)*y(6)*exp(pC*tauC)-deltaE*y(7);
ynew(8)=-betaBn*y(1)./(y(1)+hB)*y(8);
ynew(9)=betaBn*y(1)./(y(1)+hB)*y(8)*exp(pB*tauB)-deltaP*y(9);
ynew(10)=pS*y(9)-deltaS*y(10);
ynew(11)=pL*y(9)-deltaL*y(11);
ynew(12)=-betaCm*y(1)./(y(1)+hC)*y(12);
ynew(13)=betaCm*y(1)./(y(1)+hC)*y(12)*exp(pCm*tauCm)-deltaE*y(13);
\end{lstlisting}

\newpage
\thispagestyle{empty}

\begin{figure}[ht!]
\centering
\includegraphics[scale=0.6]{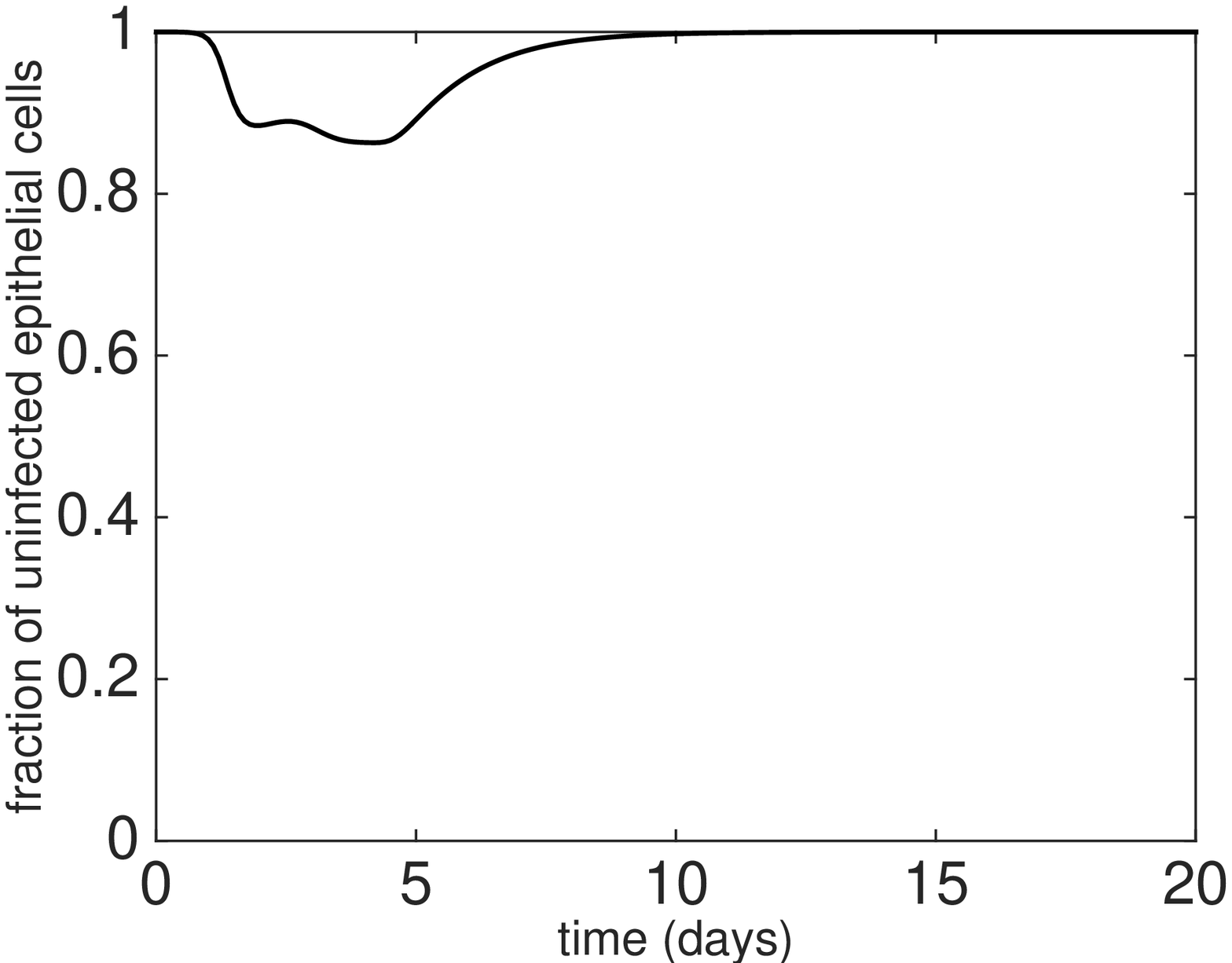}
\caption*{\small{FIGURE S1: Model solution with parameters given in Table 1 in the main text shows that the loss of uninfected epithelial cells is maintained within 10--20\% of total cell pool. This number of uninfected epithelial cells equals the sum of target cells $T$ and resistant cells $R$ shown in Fig.\ 3 in the main text.}}
\end{figure}

\newpage
\thispagestyle{empty}

\begin{figure}[ht!]
\centering
\includegraphics[scale=1]{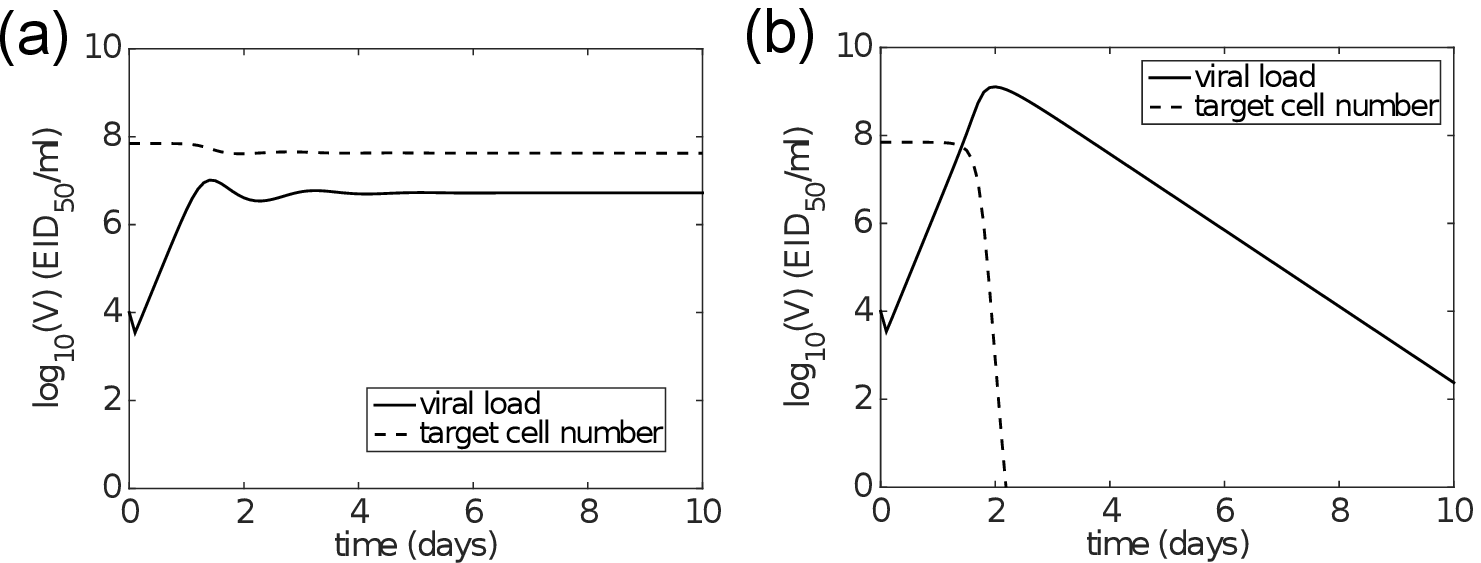}
\caption*{\small{FIGURE S2: Inclusion of IFN prevents target cell depletion. In the absence of adaptive immunity (letting $C_n=B_n=0$ in the model), the model with IFN production ($p_F = 10^{-5}$) simulates a sustained elevation of viral load and large steady state target cell number (dashed curve in panel (a)), both of which are consistent with previous experimental data (shown in Fig.\ 5 in the main text). However, in the absence of an innate response (letting $p_F = 0$), panel (b) shows that target cells are depleted, inducing a rapid fall in viral load. Note that for target cell number, the y-axis indicates log$_{10}$(target cell number).}}
\end{figure}

\newpage
\thispagestyle{empty}

\begin{figure}[ht!]
\centering
\includegraphics[scale=0.7]{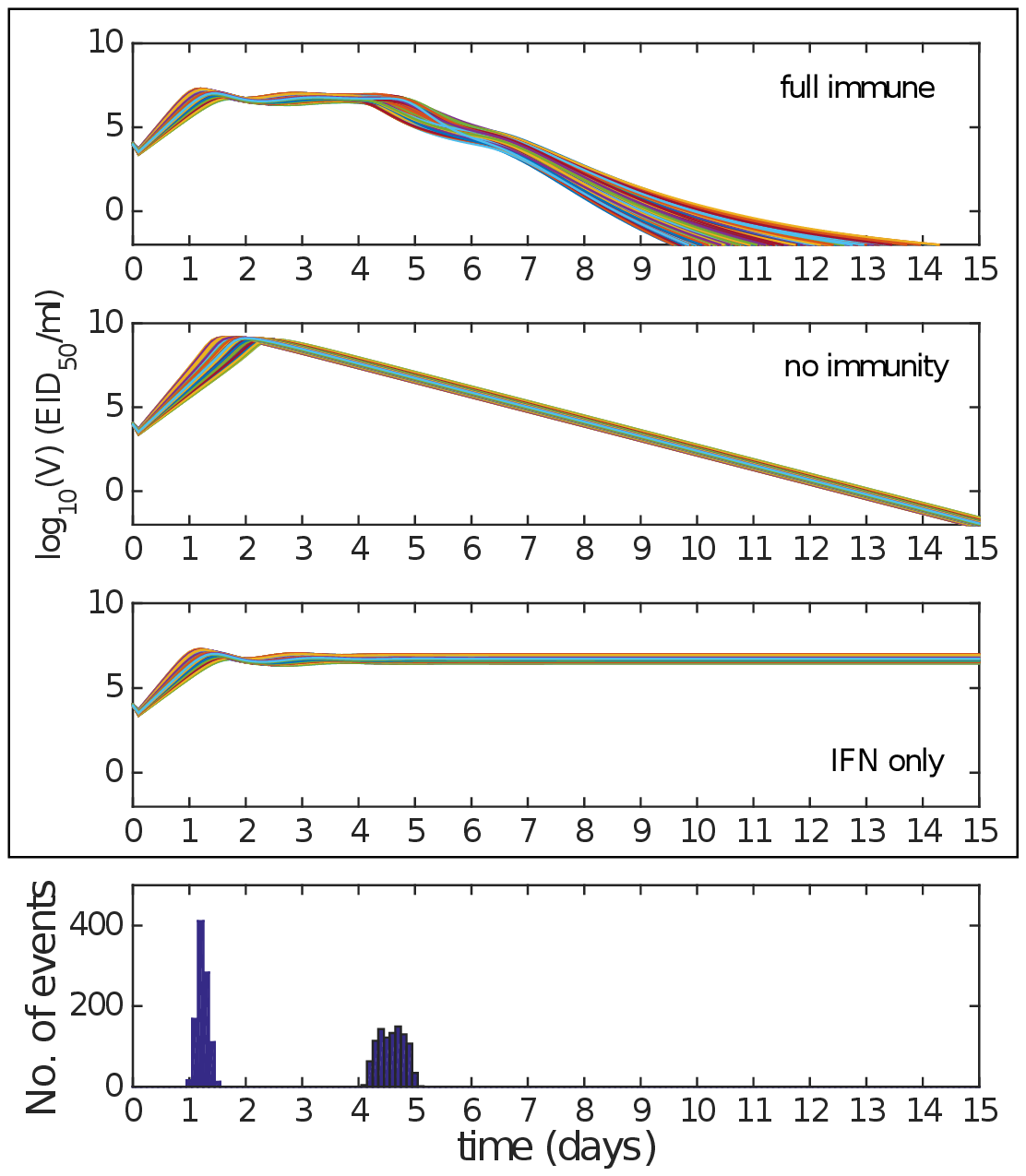}
\caption*{\small{FIGURE S3: The three-phase model behavior is robust to the change of parameters. By allowing all the parameters to vary by $\pm$20\% from their default values (i.e. those shown in Table\ 1), 1000 samples of parameter sets were selected using Latin hypercube sampling and 1000 corresponding viral load solutions are shown in the upper panel. In Fig.\ 4 in the main text, we illustrated the phase separation using area shading. Here, the rough times of phase separation are indicated by the histograms in the lower panel; the left one indicates the time separating the first and the second phases (i.e. the time when the corresponding solutions of ``full immune" and ``no immunity" differ by 0.2 (in log-scale) for the first time) and right one indicates the time separating the second and the third phases (i.e. the time when the corresponding solutions of ``full immune" and ``IFN only" differ by 0.2 (in log-scale) for the first time). The solutions of ``no immunity" are obtained by letting $p_F = 0$ and $\beta_{Cn} = \beta_{Bn} = 0$ in the model, and the solutions of ``IFN only" are obtained by letting $\beta_{Cn} = \beta_{Bn} = 0$ in the model.}}
\end{figure}

\newpage
\thispagestyle{empty}

\begin{figure}[ht!]
\centering
\includegraphics[scale=1]{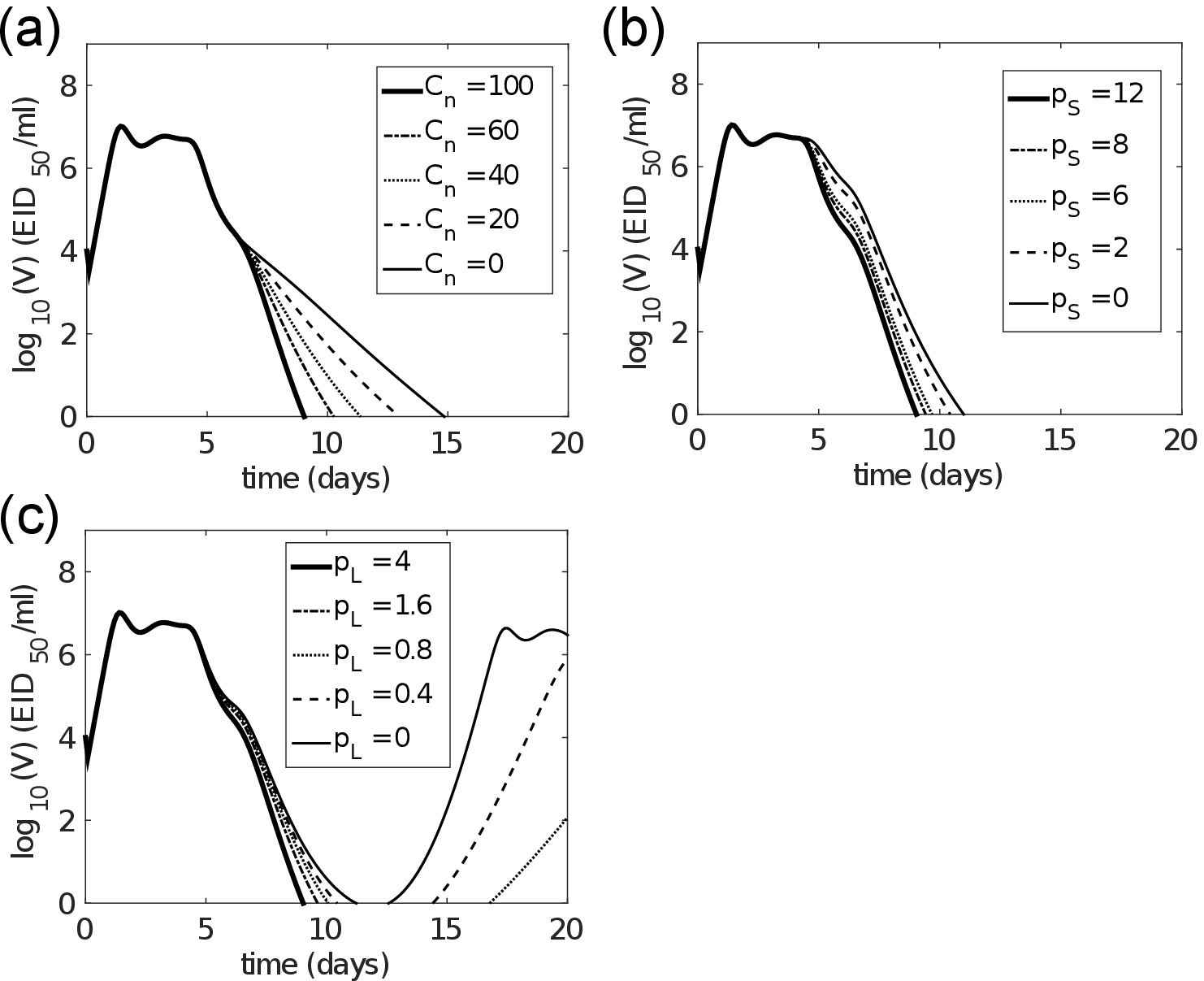}
\caption*{\small{FIGURE S4: Dependence of model behaviour on the production rate of $\cd$ T cells (panel (a)), short-lived antibodies (panel (b)) and long-lived antibodies (panel (c)). Different levels of production for $\cd$ T cells, short-lived and long-lived antibodies are modelled respectively by varying the naive $\cd$ T cell number ($C_n$), short-lived antibody production rate ($p_S$) and long-lived antibody production rate ($p_L$). For each panel, all other model parameters were kept fixed at the values given in Table 1 in the main text.}}
\end{figure}

\newpage
\thispagestyle{empty}

\begin{figure}[ht!]
\centering
\includegraphics[scale=1]{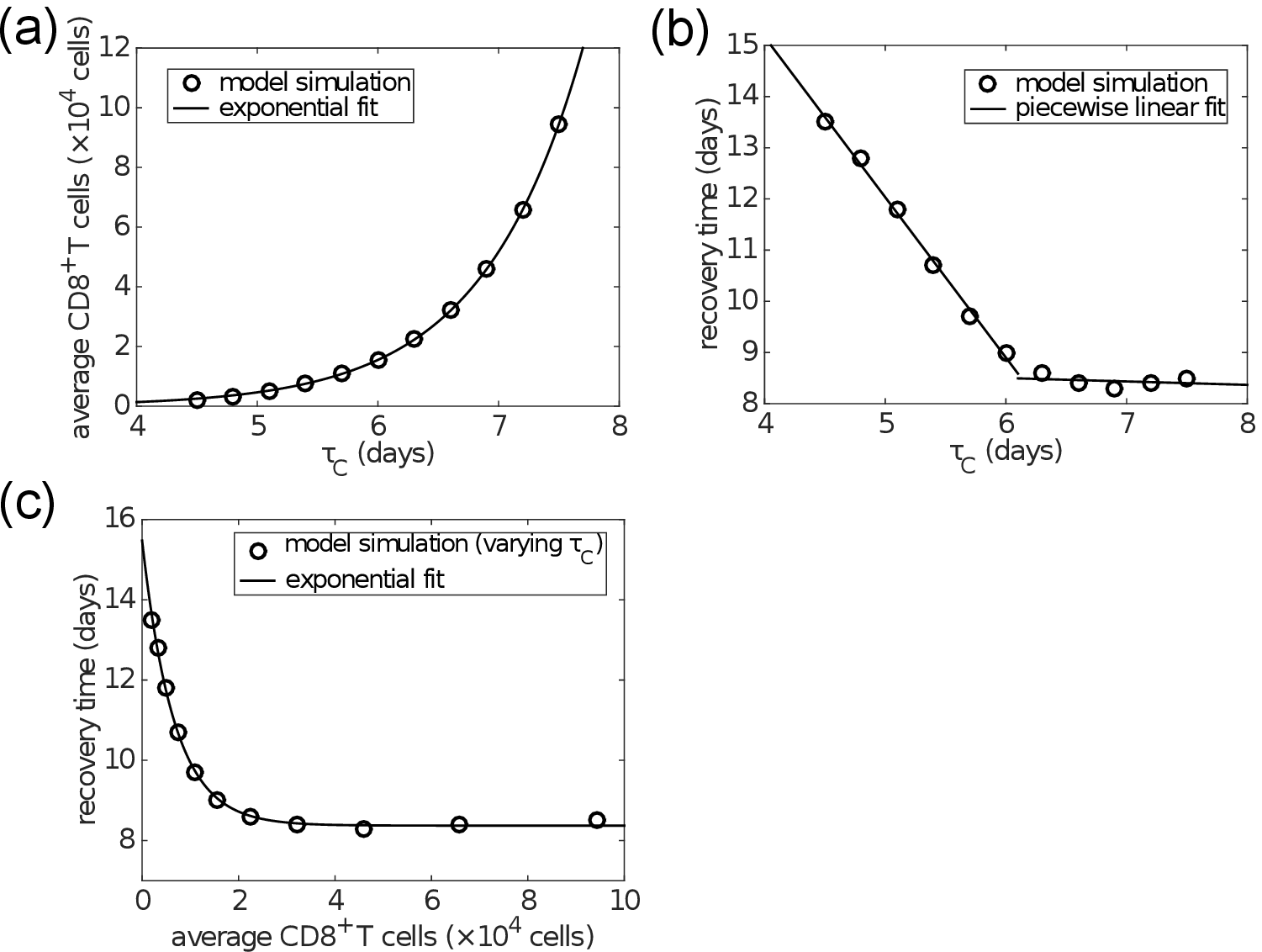}
\caption*{\small{FIGURE S5: Dependence of the average effector $\cd$ T cell number (over days 6--20) on the delay induced by naive $\cd$ T cell activation and differentiation ($\tau_C$). Recovery time is defined to be the time when viral load falls to 1 $\rm EID_{50}/ml$. Panel (a) shows that the average effector $\cd$ T cell number is exponentially related to the delay $\tau_C$. Panel (b) shows that the recovery time is related to the delay $\tau_C$ in an approximately piecewise linear manner. Panel (c) shows varying delay $\tau_C$ also preserves the approximately exponential relationship between the average $\cd$ T cell numbers and recovery time, consistent with the results of varying initial naive $\cd$ T cell numbers shown in Fig. 6 in the main text.}}
\end{figure}

\newpage
\thispagestyle{empty}

\begin{figure}[ht!]
\centering
\includegraphics[scale=0.64]{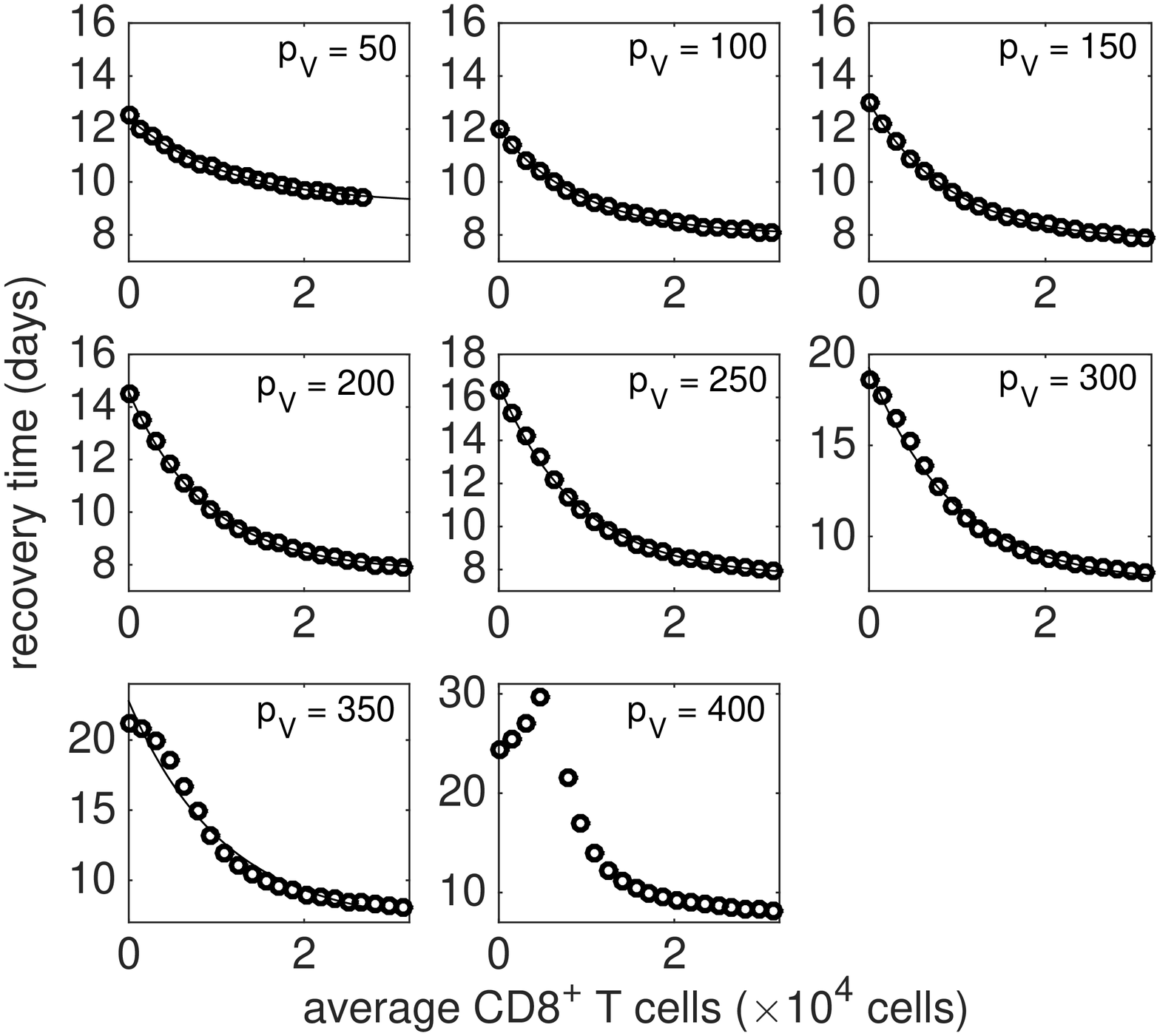}
\caption*{\small{FIGURE S6: The exponential relationship between recovery time and the average effector $\cd$ T cell number (over days 6--20) is robust to the changes in the viral production rate $p_V$. Recovery time is defined to be the time when the viral load falls to 1 $\rm EID_{50}/ml$. For each $p_V$, the relationship between average $\cd$ T cells and recovery time was obtained by varying initial naive $\cd$ T cell number.  Except for large $p_V$ (e.g. $p_V = 350, 400$), all other cases are well fit by exponential curves (thin black curves through the simulated data). When $p_V$ is large (the cases of $p_V = 350, 400$), we observe a deviation from the exponential relationship for low expression of effector $\cd$ T cells. For example a decrease in recovery time is observed for a lower level of effector $\cd$ T cells (in the range of less than 5000 cells) for $p_V = 400$, whose reason is not clear and is likely a by-product of the highly complex dynamical system.}}
\end{figure}

\newpage
\thispagestyle{empty}

\begin{figure}[ht!]
\centering
\includegraphics[scale=0.64]{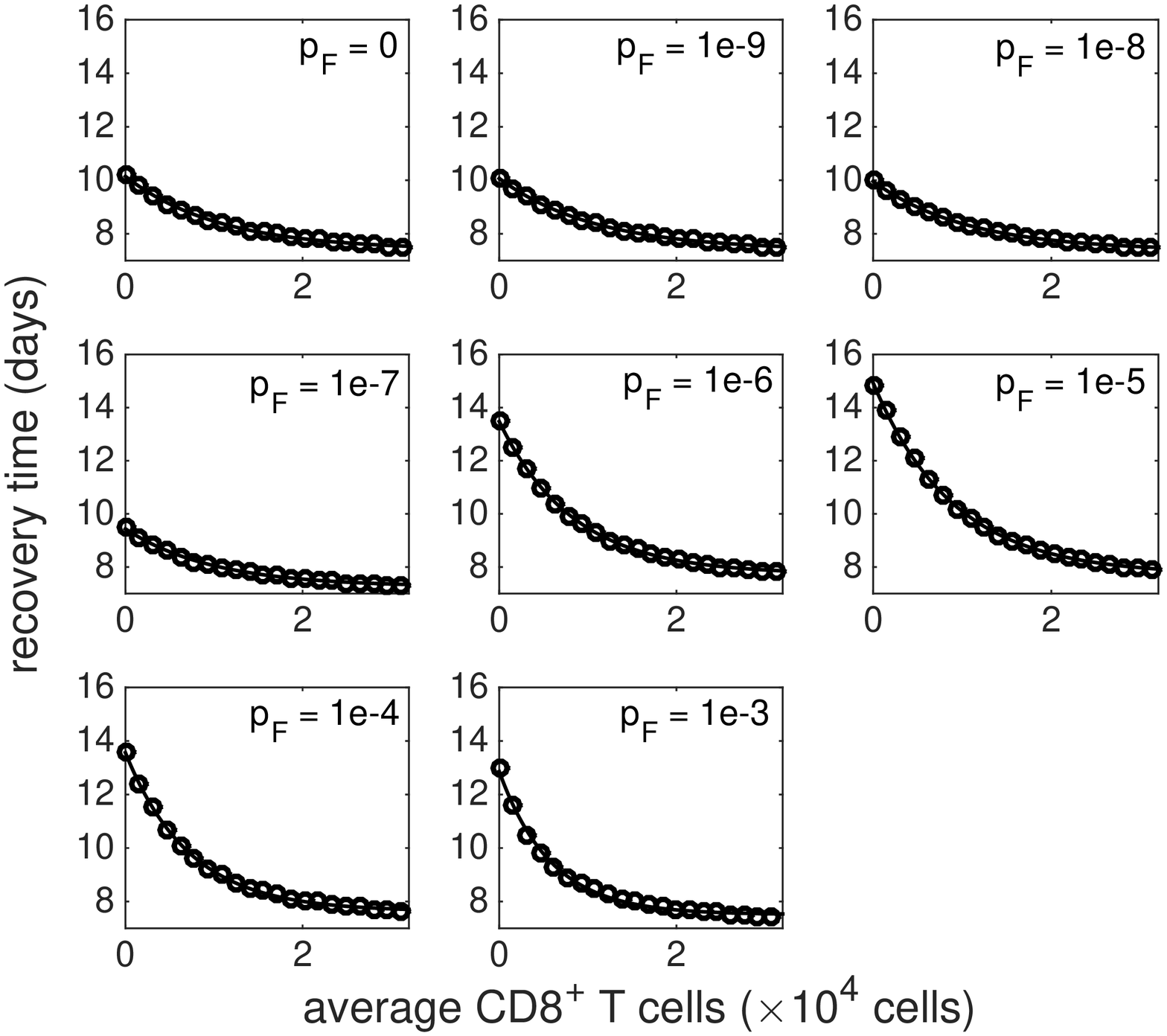}
\caption*{\small{FIGURE S7: The exponential relationship between recovery time and the average effector $\cd$ T cell number (over days 6--20) is robust to the change of IFN production rate $p_F$. Recovery time is defined to be the time when viral load falls to 1 $\rm EID_{50}/ml$. The thin black curves through the simulated data are exponential fits. One may observe that the recovery time increases and then decreases as $p_F$ increases. The reason is not entirely clear. Our explanation is that the increase for small pF is likely because increasing $p_F$ reduces the extent of target cell depletion and thus makes infection longer. When pF is sufficiently large, a further increase in $p_F$ has little effect on target cell pool but induces stronger innate immune response, which play a role in shortening the recovery time.}}
\end{figure}

\newpage
\thispagestyle{empty}

\begin{figure}[ht!]
\centering
\includegraphics[scale=0.8]{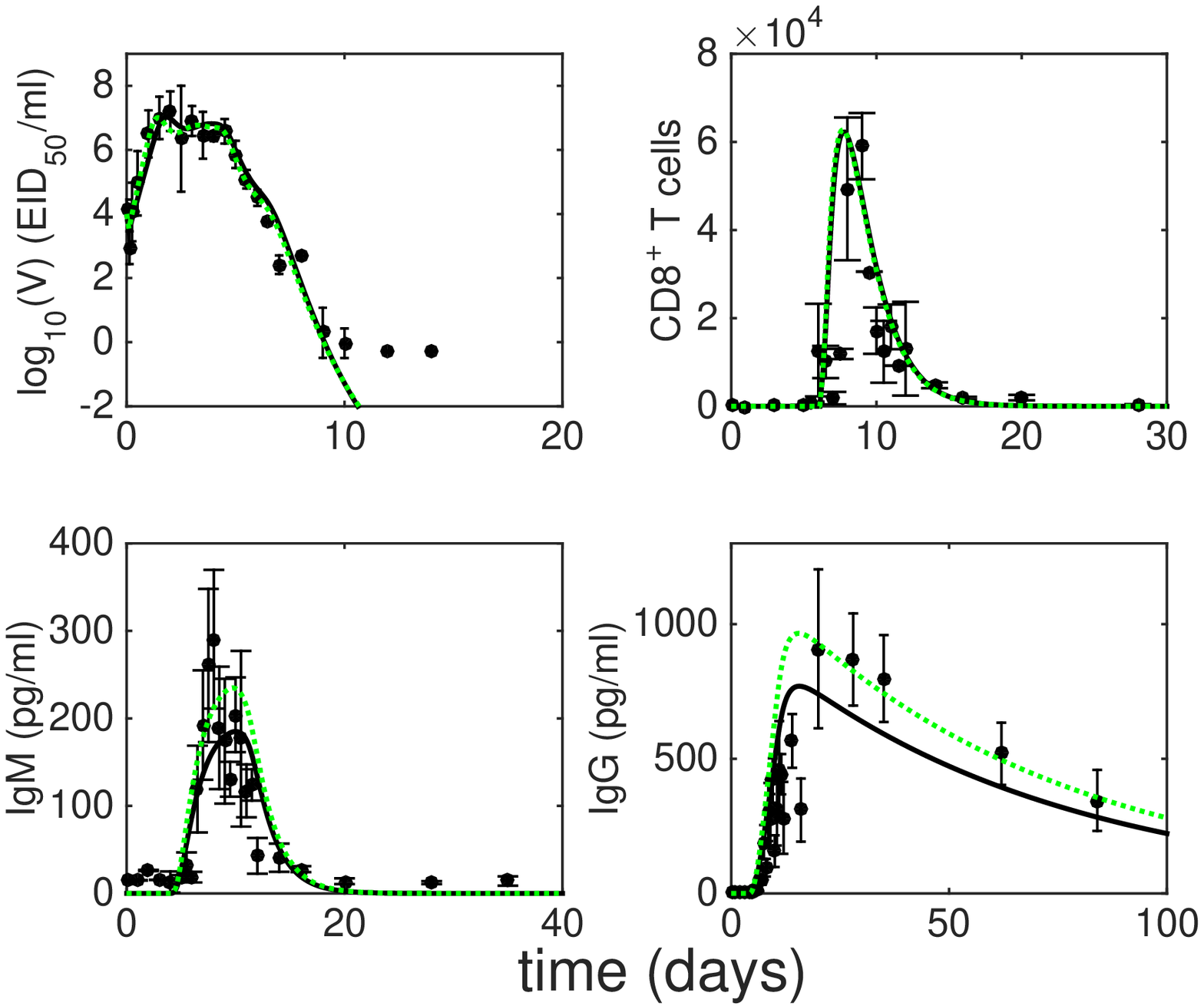}
\caption*{\small{FIGURE S8: An example of a best-fit solution failing to capture the sparse part of the IgG data very well. Black dots are experimental data from \cite{Miaoetal2010} and black curves are the fits. The fit was generated by using equally weighted LSE function (scale factors remain unchanged) and the parameters in Table 1 in the main text as an initial guess in MATLAB's \emph{fmincon} function. Compared to using weighted LSE (which generates the fit shown by green dashed curves), using equally weighted LSE does not alter the fits to the viral load and $\cd$ T cell number but significant underestimates antibody levels. Note that due to the limit of detection for the viral load (occurring after 10 days post-infection as seen in viral load data), the last three data points in the upper-left panel were not taken into consideration for model fitting.}}
\end{figure}

\newpage
\thispagestyle{empty}

\begin{figure}[ht!]
\centering
\includegraphics[scale=0.8]{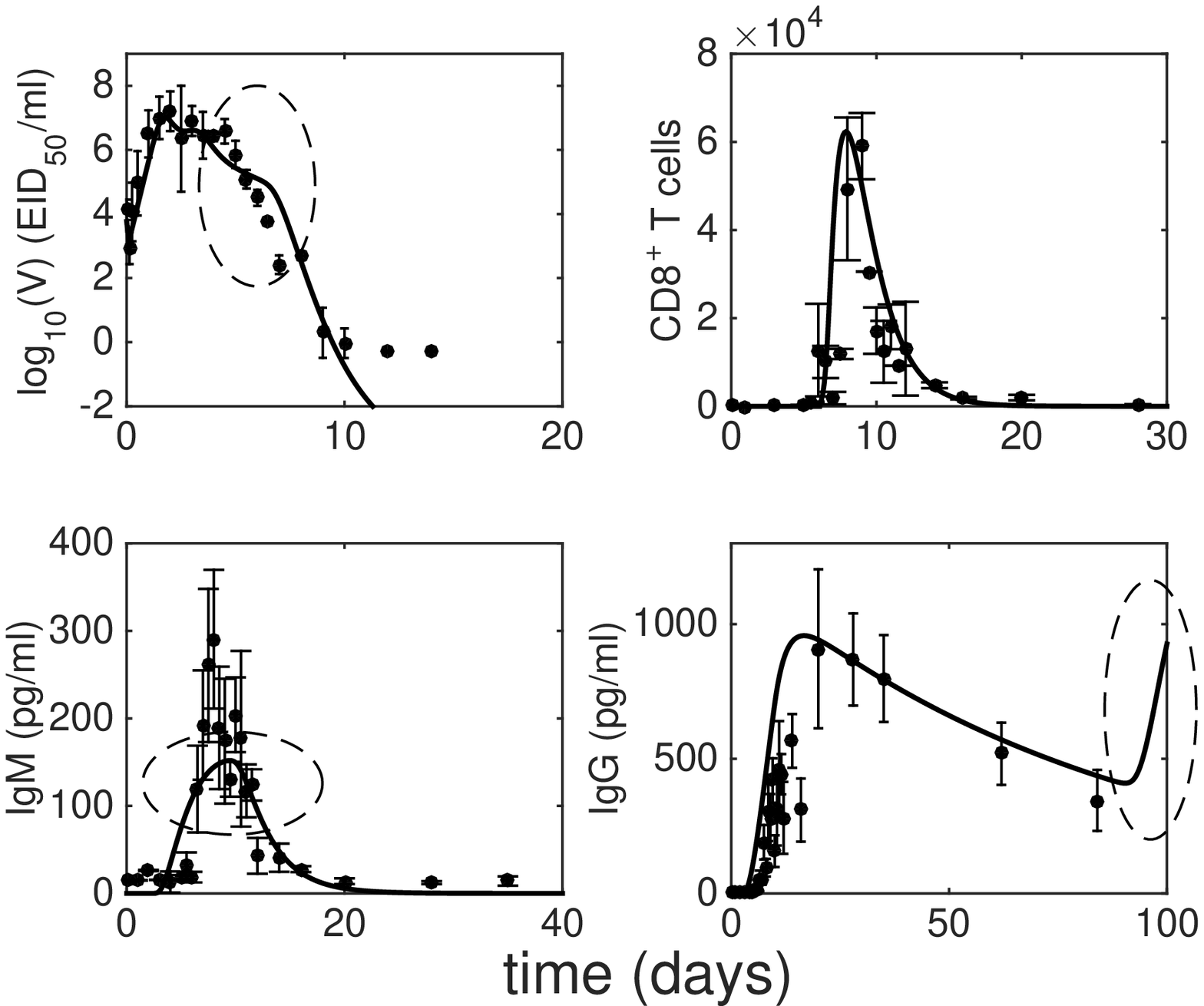}
\caption*{\small{FIGURE S9: An example of a best-fit solution exhibiting oscillatory IgG, which was deemed biologically implausible. Black dots are experimental data from \cite{Miaoetal2010} and black curves are the fits. In fact, the LSE for this fit is smaller than that generated by the estimates in Table 1 in the main text. But given the fact that a few aspects of data were not well captured by this fit (indicated by dashed ovals), we did not consider this fit as an acceptable solution. Note that due to the limit of detection for the viral load (occurring after 10 days post-infection as seen in viral load data), the last three data points in the upper-left panel were not taken into consideration for model fitting.}}
\end{figure}

\end{document}